\documentclass[aps,pre,twocolumn,groupedaddress]{revtex4-2}

\usepackage{graphicx}
\usepackage{amsmath}
\usepackage{bm}
\usepackage{amsfonts}

\begin{document}


\title{Direction selection of metachronal waves in hydrodynamic coordination of cilia}


\author{Rachel R Bennett}
\email[]{rachel.bennett@bristol.ac.uk}
\affiliation{School of Mathematics, University of Bristol, Bristol, BS8 1UG, United Kingdom}


\date{\today}

\begin{abstract}
Large arrays of active cilia coordinate their beat cycles into metachronal waves. These waves can travel in different directions with respect to the cilium's beat direction and the resulting direction of fluid propulsion. Hydrodynamic interactions provide a mechanism for the individual cilia to coordinate their beat cycles. Using an analytical framework that connects the properties of the individual cilia to the emergent wave behavior, I show how the forcing pattern of the beat cycle breaks the symmetry to select a wave direction. Previously, it was found that the second harmonic in the beat pattern is the dominant term for coordination. Here, I show that the first harmonic in the beat pattern enters the coordination dynamics at higher order and demonstrate the important role the higher order terms play in determining the direction of emerging metachronal waves.
\end{abstract}


\maketitle

\section{Introduction}
Cilia are active organelles that beat to create fluid flows for a range of tasks in biology, across many species from microorganisms to humans \cite{Gilpin:2020}.
Efficient fluid flow is created when cilia coordinate their beat cycles into metachronal waves, a form of synchronization with a phase lag between neighboring cilia \cite{Gueron:1999,Kim2006,Osterman2011,Elgeti2013,Chateau2017}. Important tasks of cilia driven flow in humans include removal of mucus from the lungs \cite{Fahy:2010,Yaghi:2016}, transporting ovulated oocytes from the ovaries \cite{Yuan:2021}, symmetry breaking in embryonic development \cite{Hirokawa:2009,Smith:2019}, and circulation of cerebrospinal fluid in brain ventricles \cite{Faubel:2016,Pellicciotta:2020}.

Recent observations of metachronal waves in natural systems include the multicellular spherical colonies, {\it Volvox} \cite{Brumley2012}; airway cells of humans \cite{Feriani:2017,Chiocciolo:2019}, other mammals and birds \cite{Burn:2022}; brain ventricle cells \cite{Pellicciotta:2020}; reef coral larvae \cite{Poon:2022} and zebrafish nose epithelia \cite{Ringers:2023}.
These results show a diversity in metachronal wave properties, such as frequency and wave direction with respect to the beat direction. Observations from older work also shows that the direction of the metachronal wave changes in different conditions: in {\it Paramecium} waves travel forward and to the right of the beat direction in normal conditions, but as the viscosity increases, both the cilium beat direction and metachronal wave direction change at different rates, and at high viscosities the metachronal wave travels backwards and to the left compared with the beat direction \cite{Machemer:1972}.
However, despite much development in experimental, theoretical and computational work, it is challenging to develop a model that can accurately predict collective cilia dynamics \cite{Cicuta:2020}, due to the multiscale nature of the problem and the diversity in structure and function in biological systems \cite{Gilpin:2020}.

Simulations and theoretical studies have shown that hydrodynamic interactions are sufficient for emergence of metachronal waves \cite{Guirao2007,Wollin2011,Elgeti2013,Brumley2015,Ghorbani2017,Han2018,Meng:2021,Westwood:2021,Solovev:2022,Kanale:2022}. Experiments on optically trapped colloids \cite{Kotar:2010,Kavre:2015,Maestro2018} and two cilia organisms \cite{Brumley2014} have confirmed that hydrodynamics are sufficient to synchronize both simplified rotor models of cilia and natural cilia.
In addition to hydrodynamic interactions, it has been demonstrated that elastic interactions via basal coupling can also provide a mechanism for cilia coordination \cite{Narematsu:2015,Wan2016,Klindt:2017,Liu2018}. Here, I address the question of how to connect wave properties to individual cilium properties when cilia are coupled through hydrodynamic interactions using a theoretical model.

Theoretical approaches to cilia coordination include 1-dimensional chains of rowers \cite{Wollin2011,Hamilton:2021,Chakrabarti:2022} and rotor models \cite{Uchida2010,Brumley2015,Meng:2021,Solovev:2022,Kanale:2022}.
Rower and rotor models represent the cilium as a sphere which moves along a prescribed trajectory, with different actuation methods in different models. Based on the typical lengthscale and beat frequency of cilia, Stokes flow is assumed so hydrodynamic interactions between the spheres are calculated using the Blake-Oseen tensor \cite{Blake1971} which includes the effects of a flat surface from which the cilia protrude.
In addition to these idealized models, filament models have been used in numerical simulations to capture the shape dynamics of the cilium \cite{Gueron1997,Kim2006,Osterman2011,Elgeti2013,Ding2014,Han2018,Schoeller:2021,Martin:2021,Westwood:2021}.

Stokes flow is time reversible, and emergence of an ordered state from a disordered state is not time reversible, so coordination can only be achieved if there are effects that can break the time-reversal symmetry. Models of two cilia systems show examples of how synchronization can be achieved, such as through a drag coefficient that varies with distance from the surface \cite{Vilfan2006}, flexibility of the cilia anchoring (which provides an additional degree of freedom) \cite{Qian2009}, or an actuation mechanism that breaks the symmetry \cite{Uchida2011,Uchida2012,Guo:2018}. These symmetry breaking features allow for the emergence of metachronal coordination in systems of many hydrodynamically coupled cilia \cite{Niedermayer2008,Uchida2010,Wollin2011,Elgeti2013,Maestro2018,Meng:2021,Kanale:2022,Chakrabarti:2022}. The symmetry breaking that occurs when a metachronal wave emerges can lead to net flow in a system of model cilia even when the individual rotors cannot generate a net flow \cite{Kanale:2022}.


In previous work, we developed a framework using a rotor model to connect the properties of individual cilia to the properties of the metachronal waves that emerge in arrays of hydrodynamically coupled model cilia \cite{Meng:2021}. This was based on the far-field hydrodynamics, but the lowest order far-field terms did not give strong enough symmetry breaking to distinguish between the stability of waves emerging in the forwards or backwards direction. A key result in this far-field model is that the dispersion relation and linear stability of emergent waves depends on the second harmonic in each rotor's driving force.
Numerical work by Hickey {\it et al.} showed that near-field effects can break the forwards-backwards symmetry \cite{Hickey:2023}. In Ref. \cite{Meng:2021}, the effect of the size of the cilium's beat cycle was neglected, but here, I develop the rotor model framework by resolving the cilium size, which changes the symmetry of interactions between cilia in different directions. I demonstrate that this symmetry breaking effect selects a direction for stable metachronal waves and introduces a strong frequency asymmetry into the dispersion relation.  
The results presented here are consistent with numerical work by Kanale {\it et al.} \cite{Kanale:2022}. They demonstrated that travelling waves emerge for both the first and second harmonic, but that there is more spatial coherence when there is a second harmonic. The analytical work presented here shows that the second harmonic term appears at lower order, indicating why the spatial coherence is stronger for this harmonic.

\section{The model}
\begin{figure}
   \includegraphics[trim=53mm 75mm 53mm 105mm,clip,width=0.9\columnwidth]{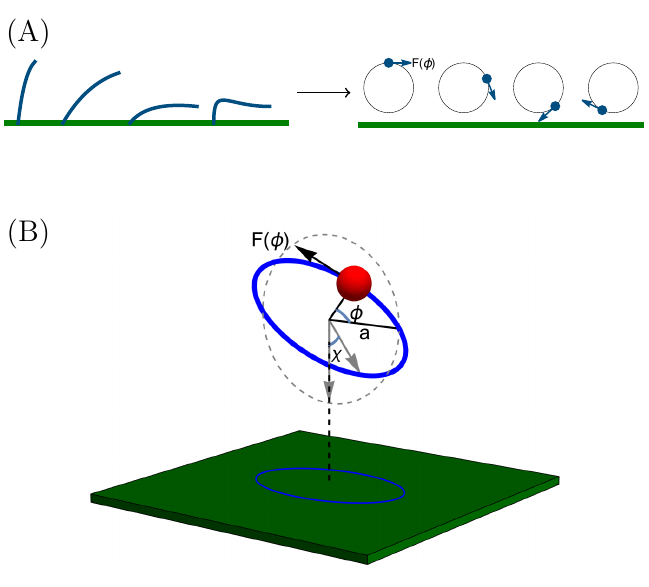}
   \caption{(A) Each cilium is modelled as a bead moving around a circular trajectory driven by a phase dependent driving force, $F(\phi)$. (B) Each trajectory has radius $a$ and is placed on a two-dimensional lattice. $\chi$ is the tilt angle of the circular trajectory; $\phi$ is the beat cycle phase.}
   \label{fig:bead_model}
\end{figure}

Following the same model set-up as in Ref. \cite{Meng:2021}, a rotor model is used here, where each cilium modelled as a bead of radius $b$ moving on a circular trajectory of radius $a$ (Fig. \ref{fig:bead_model}).
The center of the trajectory of each cilium is placed on a two-dimensional lattice with lattice position ${\bm r}$ and at height $h$ above a flat, no-slip surface. The position of the bead at lattice point ${\bm r}$ is given by
\begin{align}
   {\bm R}({\bm r},t)=&{\bm r}+h\hat{\bm z}+a(\cos{\phi({\bm r},t)}\cos{\theta} \nonumber \\
   &+\sin{\phi({\bm r},t)}\sin\theta\sin\chi,  \cos{\phi({\bm r},t)}\sin\theta \nonumber \\
   &-\sin{\phi({\bm r},t)}\cos\theta\sin\chi,h/a + \cos\chi\sin{\phi({\bm r},t)}),
\end{align}
where $\phi$ is the cilium phase, representing how far through its beat cycle it has progressed, $\theta$ is the angle the beat direction makes with the lattice, and $\chi$ is the tilt angle of the circular trajectory to the vertical.
Neighboring cilia are spaced by distance $\ell$ and the appropriate regime for this model is $a<<\ell$ and $\ell \sim h$. Each cilium bead is driven around the circular trajectory with a tangential driving force $F(\phi)$, which depends only on phase of the individual cilium. Constraint forces in the normal direction restrict the bead to stay on the circular trajectory.

The dynamical equation for a cilium on lattice point ${\bf r}$ is
\begin{align}
   \frac{d\phi({\bm r},t)}{d t}&=\frac{F(\phi({\bm r},t))}{\zeta(\phi({\bm r},t)) a}\nonumber \\
   &+\frac{1}{a}\sum_{{\bm r}'\not={\bm r}} \bm{t}(\bm{R})\cdot \bm{G}(\bm{R};\bm{R}')\cdot\bm{t}(\bm{R}') F(\phi({\bm r}',t)),
   \label{eq:dynamical}
\end{align}
where $\zeta(\phi)$ is the hydrodynamic friction (this can vary with phase due to the varying distance between the bead and the surface), $\bm{t}({\bm R})$ is tangent to the trajectory of the bead at position ${\bm R}$, and $\bm{G}(\bm{R};\bm{R'})$ is the Blake-Oseen tensor, encoding the hydrodynamic interactions felt by the bead at position $\bm{R}$ due to the force exerted by the bead at position $\bm{R}'$ \cite{Blake1971}. The first term is the intrinsic velocity of each rotor, and the second term accounts for hydrodynamic interactions with all other rotors, with the sum over all other lattice points.

The forcing and the friction are periodic and can be written as
\begin{subequations}
\begin{eqnarray}
   F(\phi) = F_{0}[1+\sum_{n=1}A_{n}\cos n\phi+B_{n}\sin n\phi], \\
   \zeta(\phi)=\zeta_{0}[1+\sum_{n=1}C_{n}\cos n\phi+D_{n}\sin n\phi].
\end{eqnarray}
\end{subequations}
The hydrodynamic friction scales as $\zeta_0=4 \pi \eta b$ and the forcing scale, $F_0$, relates to the magnitude of the forces exerted by molecular motors in a cilium filament \cite{Hill:2010}. The harmonic amplitudes, $A_n, B_n, C_n, D_n$ are small and constrained so that $F(\phi)>0$ and $\zeta(\phi)>0, \forall \phi \in \mathbb{R}$.

\section{Results}

We use this model to understand how the global emergent wave properties depend on the individual cilium parameters. In the following steps, the dynamical equation Eq. \ref{eq:dynamical} is coarse-grained and a dispersion relation is sought for emergent waves and a linear stability analysis performed (additional details of the following steps are provided in the supplementary information (SI) and a similar analysis is presented in Ref. \cite{Meng:2021}). The first step is to move to a new choice of phase coordinate, $\bar{\phi}$, so that the intrinsic velocity is constant in the new phase coordinate. The relation between the original coordinate, $\phi$, and the new coordinate $\bar{\phi}$, is
\begin{equation}
   d\bar{\phi}/d\phi=\Omega(\phi)/\Omega_0,
   \label{eq:oldphitonew}
\end{equation}
which can be integrated and expanded to linear order in the harmonic amplitudes to give
\begin{equation}
   \phi(\bar{\phi})\simeq\bar{\phi}+\sum_{n=1}\frac{1}{n}\left[\left(A_{n}-C_n\right)\sin n\bar{\phi}-\left(B_{n}-D_n\right) \cos n\bar{\phi}\right]\\
   \label{eq:relationbetweenphase}
\end{equation}
(we assume the harmonic amplitudes are small compared to one so we can neglect higher order terms in the harmonic amplitudes).

The Blake tensor is expressed using its 2-dimensional Fourier transform in the $x$ and $y$ directions, ${\bm q}=(q_x,q_y)$, and the $z$ component continues to be expressed in real space (see SI for the expressions). The dynamical equation is then expressed in the new coordinate using the Fourier space representation of the Blake tensor. Since we are interested in the long term dynamics of the emergent metachronal wave, we can use a separation of timescales between the mean phases and the phase differences, taking a time average over mean phases to simplify the dynamics. Numerical simulations in Ref. \cite{Meng:2021} show that the emergence of the wave is slow enough for this time averaging to be valid. When these steps have been performed (details in SI), the dynamical equation for $\bar{\phi}$ can be written as
\begin{eqnarray}
&&\partial_t\bar{\phi}(\bm{r}, t)=\Omega_{0}+\frac{\Omega_{0} b}{4\pi}\,\sum_{\bm{r'}} \int d^2 \bm{q} \,e^{i\bm{q}\cdot(\bm{r}-\bm{r'})} \nonumber \\
&&\times \big[({\cal M}_2(\bm{q})+\frac{a}{\ell}{\cal M}_1(\bm{q}))\cos\left({\bar{\phi}}(\bm{r})-{\bar{\phi}}(\bm{r}')\right) \nonumber \\
&&+({\cal S}_2(\bm{q})+\frac{a}{\ell}{\cal S}_1(\bm{q}))\sin\left({\bar{\phi}}(\bm{r})-{\bar{\phi}}(\bm{r}')\right), \nonumber \\
&& +\frac{a}{\ell}K_1({\bm q})\cos\left(2({\bar{\phi}}(\bm{r})-{\bar{\phi}}(\bm{r}'))\right) \nonumber \\
&&+\frac{a}{\ell}J_1({\bm q})\sin\left(2({\bar{\phi}}(\bm{r})-{\bar{\phi}}(\bm{r}'))\right)+\frac{a}{\ell}P_1(\bm q)   \big]
\label{eq:dynamicalnewphi}
\end{eqnarray} 
${\cal M}_2$ and ${\cal S}_2$ contain the second harmonic amplitudes ($A_2, B_2, C_2, D_2$, representing a nematic beat pattern) and geometric parameters $\theta, \chi, h$. ${\cal M}_1, {\cal S}_1, K_1, J_1, P_1$ contain the first harmonic amplitudes ($A_1, B_1, C_1, D_1$, representing a polar beat pattern) and geometric parameters $\theta, \chi, h$. Higher harmonic amplitudes do not appear at the order expanded to here. Previously in Ref. \cite{Meng:2021}, we expanded the hydrodynamic interaction term to leading order in the trajectory size, and therefore the only harmonics obtained were second harmonics, which have $\phi \to \phi+\pi$ symmetry. In Eq. \ref{eq:dynamicalnewphi}, I expand to first order in the trajectory size and the higher order terms that are obtained contain the first harmonic that do not have $\phi \to \phi+\pi$ symmetry. Expressions for all coefficients in Eq. \ref{eq:dynamicalnewphi} are given in the SI.

\subsection{Dispersion relation}
We seek a dispersion relation for metachronal waves. A metachronal wave has the form
\begin{equation}
   \bar{\phi}({\bm r},t)=\omega t-{\bm k}\cdot {\bm r}+\delta\bar{\phi}_{\bm k}({\bm r},t),
   \label{eq:waveansatz}
\end{equation}
where $\delta\bar{\phi}_{\bm k}$ is a perturbation to the wave of wavevector ${\bm k}$. Eq. \ref{eq:waveansatz} is inserted into Eq. \ref{eq:dynamicalnewphi} and expanded to first order in the perturbation $\delta\bar{\phi}_{\bm k}$. The dispersion relation is obtained from the zeroth order terms and the linear stability equation is obtained from the first order terms.
The  zeroth order terms give the following dispersion relation:
\begin{align}
   \omega({\bm k})=\Omega_0+\Omega_0\frac{\pi b}{\ell^2}\sum_{\bm G}\Bigg[\frac{M({\bm k}+{\bm G})+M(-{\bm k}+{\bm G})}{2}\nonumber \\
   +\frac{S({\bm k}+{\bm G})-S(-{\bm k}+{\bm G})}{2{\rm i}}\\
   +\frac{a}{\ell}\frac{K_1(2{\bm k}+{\bm G})+K_1(-2{\bm k}+{\bm G})}{2} \nonumber\\
   +\frac{a}{\ell}\frac{J_1(2{\bm k}+{\bm G})-J_1(-2{\bm k}+{\bm G})}{2{\rm i}}+\frac{a}{\ell}P_1({\bm G})\Bigg],
\end{align}
where the sum is over the reciprocal lattice vectors, and $M({\bm q})={\cal M}_2({\bm q})+\frac{a}{\ell}{\cal M}_1({\bm q}), S({\bm q})={\cal S}_2({\bm q})+\frac{a}{\ell}{\cal S}_1({\bm q})$. 
In contrast to the lowest order calculation in Ref. \cite{Meng:2021}, where only second harmonic terms appear in the dispersion relation, we see that with the additional terms that account for trajectory size, first harmonic symmetry breaking terms appear in the dispersion relation.

\subsection{Linear stability analysis}
Expanding Eq. \ref{eq:dynamicalnewphi} with the metachronal wave form (Eq. \ref{eq:waveansatz}) to linear order in the perturbation gives
\begin{equation}
   \partial_t \delta\bar{\phi}_{\bm k}({\bm r},t)=\sum_{{\bm r}'}\gamma_{\bm k}({\bm r}-{\bm r}')(\delta\bar{\phi}_{\bm k}({\bm r},t)-\delta\bar{\phi}_{\bm k}({\bm r}',t)).
   \label{eq:perturbwavereal}
\end{equation}
Taking the discrete Fourier transform of Eq. \ref{eq:perturbwavereal} gives us
\begin{equation}
   \partial_t \delta\tilde{\phi}_{\bm k}({\bm q}',t)=(\tilde{\gamma}_{\bm k}(0)-\tilde{\gamma}_{\bm k}({\bm q}'))\delta\tilde{\phi}_{\bm k}({\bm q}',t),
   \label{eq:perturbwavereciprocal}
\end{equation}
where $\delta\tilde{\phi}_{\bm k}({\bm q}',t)$ is the two-dimensional discrete Fourier transform of $\delta\bar{\phi}_{\bm k}({\bm r},t)$ and $\tilde{\gamma}_k({\bm q}')$ is given by
\begin{align}
   \tilde{\gamma}_{\bm k}({\bm q'})=\Omega_0\frac{\pi b}{\ell}\sum_{\bm G}\Bigg[\frac{-M(-{\bm q}'+{\bm k}+{\bm G})+M(-{\bm q}'-{\bm k}+{\bm G})}{2{\rm i}}\nonumber \\
   +\frac{S(-{\bm q}'+{\bm k}+{\bm G})+S(-{\bm q}'-{\bm k}+{\bm G})}{2}\nonumber \\
   +\frac{a}{\ell}\frac{-K_1(-{\bm q}'+2{\bm k}+{\bm G})+K_1(-{\bm q}'-2{\bm k}+{\bm G})}{{\rm i}} \nonumber \\
   +\frac{a}{\ell}\big(J_1(-{\bm q}'+2{\bm k}+{\bm G})+J_1(-{\bm q}'-2{\bm k}+{\bm G})\big)   \Bigg].
\end{align}
For a metachronal wave with wavevector ${\bm k}$ to be linearly stable, we require $(\tilde{\gamma}_{\bm k}(0)-\tilde{\gamma}_{\bm k}({\bm q}'))<0$ for all perturbation wavevectors ${\bm q}'$. Again, we see that both the first and second harmonic terms appear in $\tilde{\gamma}_{\bm k}({\bm q}')$, so the first harmonic can break the symmetry between the stability of forwards and backwards waves.

\begin{figure*}
\centering
   \begin{tabular}{lll}
   (A) & (B) & (C) \\
   \includegraphics[width=0.3\textwidth]{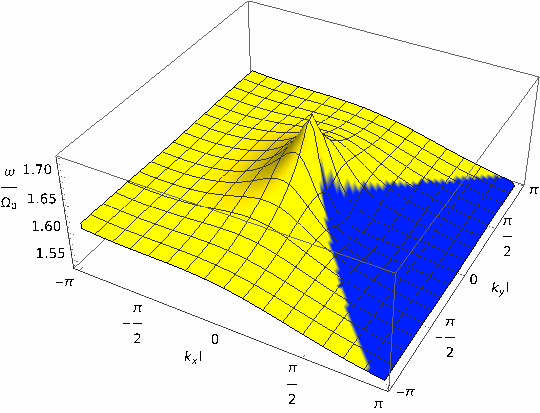} 
   &
   \includegraphics[width=0.3\textwidth]{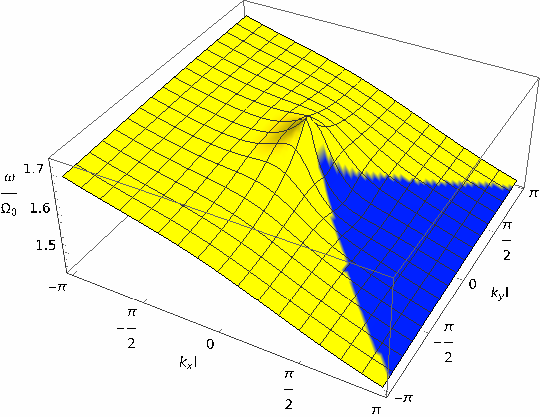} 
   &
   \includegraphics[width=0.3\textwidth]{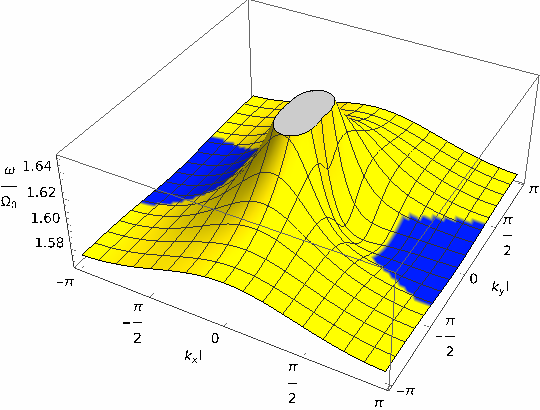} 
   \\
   (D) & (E) & (F) \\
   \includegraphics[width=0.3\textwidth]{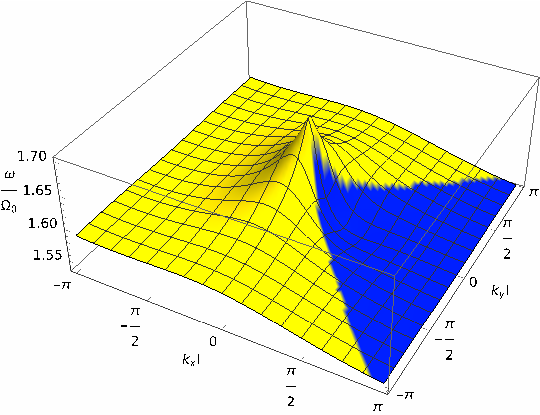} 
   &
   \includegraphics[width=0.3\textwidth]{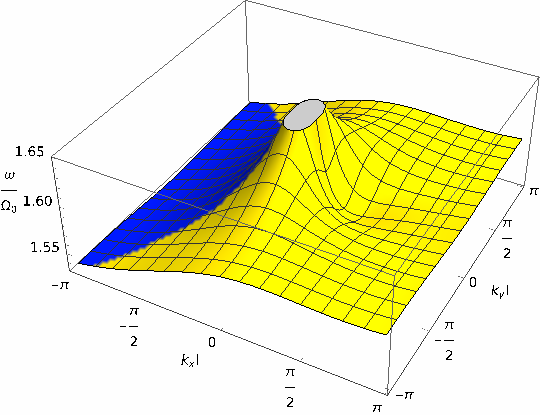} 
   &
   \includegraphics[width=0.3\textwidth]{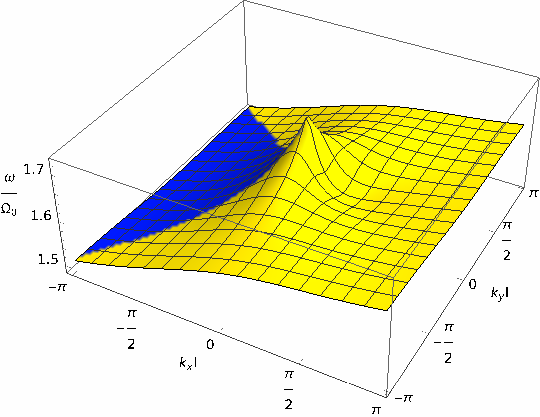} \\
   \end{tabular}
   \caption{Dispersion relation of metachronal waves using a square lattice. Linear stability analysis is represented by color: yellow regions are unstable and blue regions are stable. Parameters used for all plots are $a/\ell=0.3$, $b/\ell=0.03$, $h=\ell$. Parameters for individuals plots are: (A) $\theta=0, \chi=0, A_1=0.1, B_1=0.04, A_2=0.1, B_2=0.05, D_1=-0.1, C_1=C_2=D_2=0$; (B) $\theta=0.01, \chi=0.1, A_1=0.3, B_1=0.4, A_2=0.1, B_2=0.05, D_1=-0.1, C_1=C_2=D_2=0$; (C) $\theta=0, \chi=0, A_1=B_1=0, A_2=0.1, B_2=0.05, C_1=D_1=C_2=D_2=0$ (second harmonic only, full range not shown for small $|{\bm k}|$); (D) $\theta=0, \chi=0, A_1=0.1, B_1=0.1, A_2=B_2=0, C_1=D_1=C_2=D_2=0$ (first harmonic only); (E) $\theta=0, \chi=0, A_1=-0.1, B_1=-0.1, A_2=B_2=0, C_1=D_1=C_2=D_2=0$ (first harmonic only, opposite sign, full range not shown for small $|{\bm k}|$); (F) $\theta=0.01, \chi=0.1, A_1=-0.3, B_1=-0.4, A_2=0.1, B_2=0.05, D_1=-0.1, C_1=C_2=D_2=0$.}
   \label{fig:results_square}
\end{figure*}

\begin{figure*} 
   \centering
   \begin{tabular}{lll}
   (A) & (B) & (C) \\
   \includegraphics[width=0.3\textwidth]{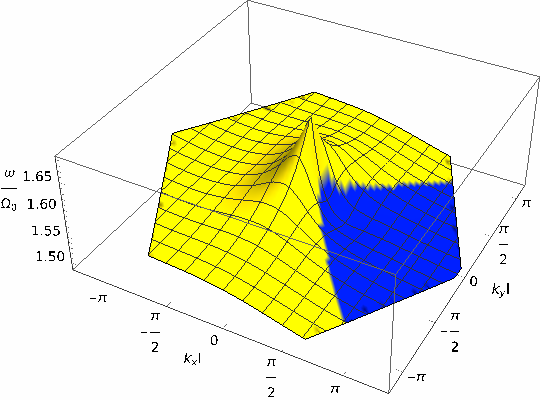} 
   &
   \includegraphics[width=0.3\textwidth]{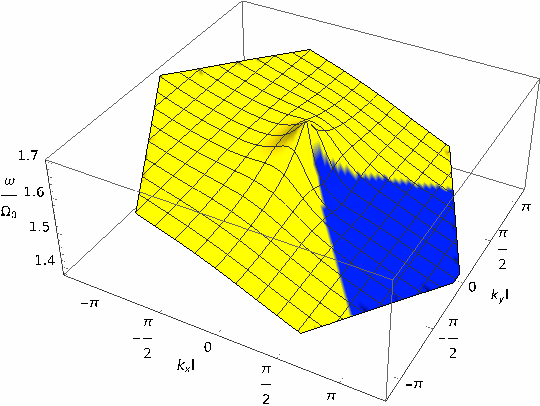} 
   &
   \includegraphics[width=0.3\textwidth]{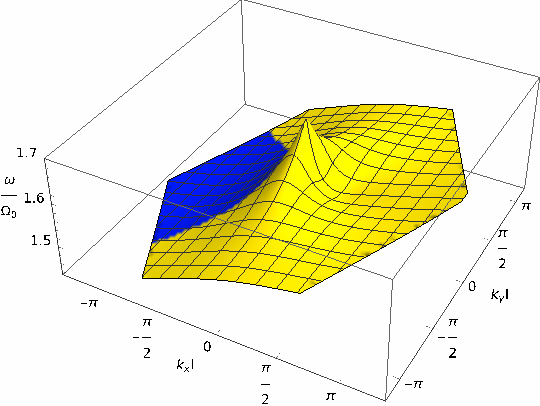} 
   \\
   (D) & (E) & (F) \\
   \includegraphics[width=0.3\textwidth]{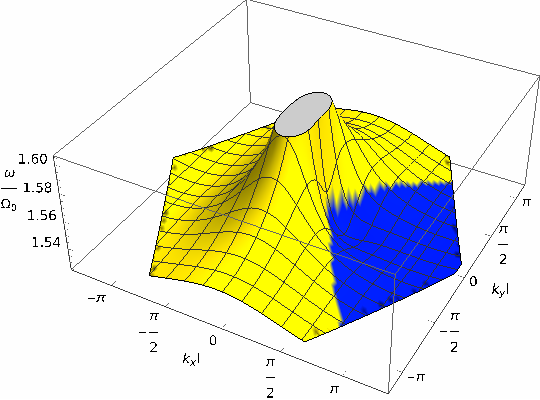} 
   &
   \includegraphics[width=0.3\textwidth]{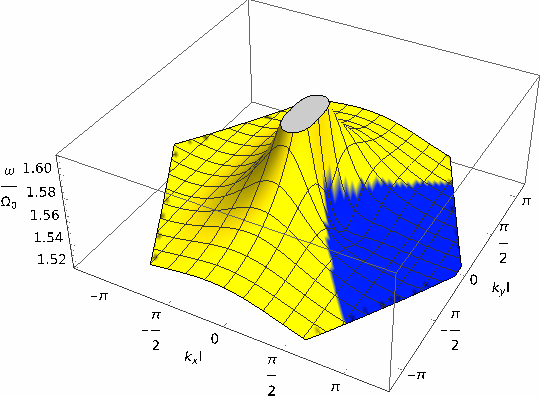}
   &
   \includegraphics[width=0.3\textwidth]{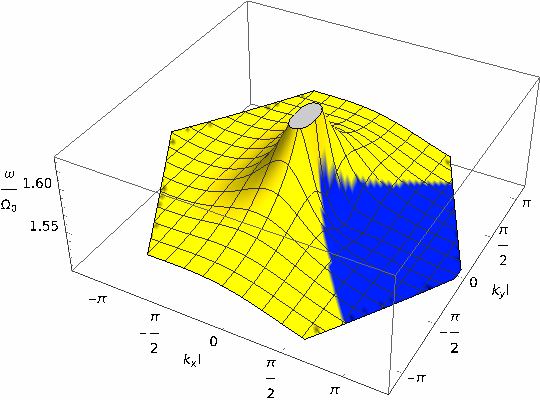} \\
   \end{tabular}
   \caption{Dispersion relation of metachronal waves using a hexagonal lattice. Linear stability analysis is represented by color: yellow regions are unstable and blue regions are stable. Parameters used for all plots are $b/\ell=0.03$, $h=\ell$. Parameters for individuals plots are: (A) $a/\ell=0.3$, $\theta=0, \chi=0, A_1=0.1, B_1=0.04, A_2=0.1, B_2=0.05, D_1=-0.1, C_1=C_2=D_2=0$; (B) $a/\ell=0.3$, $\theta=0.01, \chi=0.1, A_1=0.3, B_1=0.4, A_2=0.1, B_2=0.05, D_1=-0.1, C_1=C_2=D_2=0$; (C) $a/\ell=0.3$, $\theta=0.01, \chi=0.1, A_1=-0.3, B_1=-0.4, A_2=0.1, B_2=0.05, D_1=-0.1, C_1=C_2=D_2=0$; (D) $a/\ell=0.04$, $\theta=0, \chi=0, A_1=0.3, B_1=0.4, A_2=0.1, B_2=0.05, D_1=-0.1, C_1=C_2=D_2=0$; (E) $a/\ell=0.1$, $\theta=0, \chi=0, A_1=0.3, B_1=0.4, A_2=0.1, B_2=0.05, D_1=-0.1, C_1=C_2=D_2=0$; (F) $a/\ell=0.2$, $\theta=0, \chi=0, A_1=0.3, B_1=0.4, A_2=0.1, B_2=0.05, D_1=-0.1, C_1=C_2=D_2=0$. (Full range not shown for small $|{\bm k}|$ for (D),(E),(F).)}
   \label{fig:results_hex}
\end{figure*}

In Fig. \ref{fig:results_square} the dispersion relation and regions of linear stability are shown for different choices of harmonic amplitudes on a square lattice. In Fig. \ref{fig:results_hex} results are shown for a hexagonal lattice. We see that when there is only a second harmonic, there is a stable region in both the forwards and the backwards direction (Fig. \ref{fig:results_square}(C)). However, when there is a first harmonic, the stable region is restricted to either the forwards or backwards direction, depending on the sign of the first harmonic amplitudes. 
There are no significant differences between the square lattice and hexagonal lattice. Additional dispersion and stability plots are shown in the SI in FIg. S1 using other parameter combinations. These results show that although the symmetry breaking effects occur at higher order, they are strong enough to restrict the stable region to only the forwards or only the backwards direction.

\section{Discussion}
Resolving the size of the cilium beat trajectory in this framework significantly affects the symmetry of the system. The symmetry breaking introduced through trajectory size is necessary to determine the direction of stable metachronal waves, which depends on the first harmonic amplitudes. In previous work \cite{Meng:2021}, when the trajectory size was neglected, the first harmonic amplitude did not appear in the linear stability calculation or the dispersion relation.
For harmonic coefficients $A_1, B_1>0$ in the force pattern, the stable region is found in the forwards direction, and for $A_1, B_1<0$ the stable region is found in the backwards direction. When $A_1$ and $B_1$ have opposite signs, the stability region changes shape (see SI Fig. S1). Further details of the effects of the sign in the first harmonic amplitudes is shown in Fig. S1 of the SI, showing that both $A_1$ and $B_1$ influence the stable direction.
With a first harmonic in the force pattern, the stable region is wider and expands to smaller wavevectors ${\bm k}$ than each stable region when there is only a second harmonic (Fig. \ref{fig:results_square}E). Reducing the size of the trajectory or magnitude of the first harmonic coefficients reduces the stable region only slightly, eliminating the smallest wavenumbers (this framework uses an infinite lattice, so there is no lower bound on the wavenumber; finite size effects would introduce a lower bound).

Waves in different direction are seen in nature: e.g. {\it Volvox carteri} shows symplectic metachronal coordination (waves travelling in the same direction as the beat direction) \cite{Brumley2012}, {\it Paramecium} shows dexioplectic waves (waves travelling to the right of the beat direction) in normal conditions, but direction changes are observed with increasing viscosity \cite{Machemer:1972}, and rabbit trachea show antilaeoplectic waves (wave direction backwards and to the left compared with beat direction) \cite{Sanderson:1981}. The work presented here shows how different metachronal wave directions can emerge due stability changes with different individual cilium beat patterns. Previous work using different modelling approaches has shown that antiplectic waves, which travel in the direction opposite to the beat direction, are more efficient than symplectic waves \cite{Gauger2009,Osterman2011,Khaderi2011,Ding2014,Chateau2017,Chateau:2019}. 
Other wave directions seen in nature suggest that efficiency is not the only criteria for the evolution of a particular organ or organism's wave direction; Guo {\it et al.} showed that this is also the case for the kinetics of the individual cilium beat pattern \cite{Guo:2014}. The work here shows that not all metachronal wave directions are stable, so wave stability is a factor in determining direction.

The dispersion relation shows a peak at ${\bm k}=0$, but also shows symmetry breaking between waves in the forward direction and waves in the backwards direction. Higher frequencies are observed at smaller wavenumbers when neighboring cilia are close in phase, so the fluid flow created by a cilium helps its nearest neighbors to move around their trajectories faster.
There is a frequency asymmetry between waves in the forwards and backwards direction that depends on $B_1$. The frequency asymmetry between the forwards and backwards direction becomes stronger as the trajectory size increases or as $B_1$ increases. Increasing the size of the trajectory or the magnitude of the first harmonic coefficients increases the wave frequency $\omega$ across the range of ${\bm k}$ in the first Brillouin zone. In most cases, the stable direction corresponds to the direction with lower wave frequency, although there is a counter example of this in the SI in Fig. S1B where $A_1$ and $B_1$ have opposite signs.

Symmetry breaking in the leading order term is very weak and is not sufficient to select a preferred direction for stable metachronal waves. There is a very weak symmetry breaking effect due to the term in ${\cal S}_2({\bm q})$ that is linear in $q_\alpha$ (see SI Eq. S14). 
This weak effect gives a small change in the shape of the dispersion relation between the forwards and backwards direction, but is not present in the linear stability analysis.
This weak symmetry breaking is caused by terms in the Blake tensor that couples the forces and velocities between the directions perpendicular and parallel to the surface in a non-reciprocal manner.
By introducing the trajectory size at linear order, there is a non-reciprocity between the interaction with the neighbor in front and the neighbor behind, due to the difference in distances. This gives a much stronger symmetry breaking effect that appears in both the dispersion relation and the linear stability analysis, leading to direction selection of stable travelling waves.

The effects of higher harmonics are negligible; only terms that are linear in the harmonic amplitude are considered in this model and higher order harmonics would only appear at higher order combinations of harmonic amplitudes. 

Work by Hickey {\it et al.} investigates numerically how coordination develops in a rotor system and show that coordinated behavior is seeded at the edges of a finite system and spreads across the system in a time that is linearly dependent on the system size \cite{Hickey:2023}. It would be interesting in future work to develop these analytical tools to study finite systems and edge effects. Another important direction for future study is developing a theoretical framework for understanding elastic coordination.

\section{Conclusion}
Here, I have presented a framework that connects the properties of an individual cilium to the properties of the emergent metachronal waves, namely the dispersion relation and the global stability. This work develops the framework initially presented in Ref. \cite{Meng:2021}, which only includes far-field hydrodynamic terms that do not account for the size of the individual cilium's beat trajectory. These far-field terms show very little difference between waves in the forwards and backwards direction, which can be seen by the presence of only the second harmonic. Here, I have expanded the hydrodynamic interactions to first order in the trajectory size, which introduces the first harmonic in the beat pattern that provides strong symmetry breaking between the forwards and backwards direction. This framework shows that although the second harmonic of the beat pattern is the leading order term for coordination, the first harmonic can also lead to emergence of global metachronal waves, significantly affect the region of linear stability, and is necessary for selecting a stable direction of metachronal waves.

\begin{acknowledgments}
R.R.B. is grateful for fruitful discussions with Tanniemola Liverpool, Alexander Mietke and Ramin Golestanian and acknowledges funding from a University of Bristol Vice-Chancellor's fellowship.
\end{acknowledgments}



%

\onecolumngrid
\newpage

\begin{center}
  \textbf{\large Direction selection of metachronal waves in hydrodynamic coordination of cilia:\\ Supplementary Information}\\[.2cm]
  Rachel R. Bennett\\[.1cm]
  {\itshape School of Mathematics, University of Bristol, Bristol, BS8 1UG, United Kingdom\\}
  rachel.bennett@bristol.ac.uk\\
\end{center}

\setcounter{equation}{0}
\setcounter{section}{0}
\setcounter{figure}{0}
\setcounter{table}{0}
\setcounter{page}{1}
\renewcommand{\theequation}{S\arabic{equation}}
\renewcommand{\thefigure}{S\arabic{figure}}
\renewcommand{\bibnumfmt}[1]{[S#1]}
\renewcommand{\citenumfont}[1]{S#1}
\allowdisplaybreaks[1]

The dynamical equation for rotor at lattice site ${\bm r}$ on a fixed trajectory with phase $\phi({\bm r})$, tangent $\hat{\bm t}(\phi({\bm r}))$, being driven by a driving force ${\bm F}=F(\phi({\bm r}))\hat{\bm t}(\phi(\bm r))$ and experiencing hydrodynamic friction $\zeta(\phi({\bm r}))$ is
\begin{equation}
   \frac{d\phi({\bm r})}{d t}=\frac{F(\phi(\bm r))}{\zeta(\phi(\bm r)) a}+\frac{1}{a}\sum_{{\bm r}'\not={\bm r}} \bm{t}(\phi({\bm r}))\cdot \bm{G}(\bm{R}({\bm r});\bm{R}({\bm r}'))\cdot\bm{t}(\phi({\bm r}')) F(\phi({\bm r}')),
   \label{eq:dynamical}
\end{equation}

We are interested in rotors above a plane surface, so we use the Blake tensor for hydrodynamic interactions between pairs of rotors \cite{Blake1971}.

\section{Blake tensor for hydrodynamic interactions}
\subsection{The Blake tensor in real space}
The Blake tensor, ${\bm G}({\bm R};{\bm R}')$, is used to calculate the flow at point ${\bm R}$ due to a point force acting at point ${\bm R}'=(x',y',z')$ above a plane, no-slip surface. It is constructed using the solution for a point force in an infinite domain, and then adding image singularities at point $\bar{\bm R}'=(x',y',-z')$ to satisfy the boundary conditions on the surface. In real space, the Blake tensor is
\begin{eqnarray}\label{Blake1}
G_{ij}({\bm R};{\bm R}')&=&\frac{1}{8\pi\eta}\left\{\frac{\delta_{ij}}{d}+\frac{d_{i}d_{j}}{d^{3}}-
\frac{\delta_{ij}}{\bar{d}}-\frac{\bar{d}_{i}\bar{d}_{j}}{\bar{d}^{3}}+
2z(\delta_{j\alpha}\delta_{\alpha\xi}-\delta_{j 3}\delta_{3\xi})\frac{\partial}{\partial \bar{d}_{\xi}}
\left[\frac{z\bar{d}_{i}}{\bar{d}^{3}}-
\left(\frac{\delta_{i 3}}{\bar{d}}+\frac{\bar{d}_{i}\bar{d}_{3}}{\bar{d}^{3}}
\right)\right]\right\},
\end{eqnarray}
with $\bm{d}=\bm{R}-\bm{R}'$, $\bm{\bar{d}}=\bm{R}-\bar{\bm R}'$. Greek letters denote the directions in 3D space with summation convention. 

\subsection{Two-dimensional Fourier transform of the Blake tensor}
Blake tensor can be expressed in two-dimensional Fourier space $\bm{q}=(q_{x},q_{y})$. The projection of ${\bm R}$ and ${\bm R}'$ into the $x-y$ plane is denoted as ${\bm R}_p=(x,y)$ and ${\bm R}_p'=(x',y')$. Using the Fourier transforms of the components of the Blake tensor, we can write

\begin{eqnarray}
G_{\alpha\beta}({\bm R};{\bm R}')&=&\frac{1}{16\pi^{2}\eta}\int d^{2}q\frac{1}{|\bm{q}|}e^{i\bm{q}\cdot(\bm{R}_{p}-\bm{R}_{p}')}\left\{\left(2\delta_{\alpha\beta}-\frac{q_{\alpha}q_{\beta}}{|\bm{q}|^{2}}\right)
\left[e^{-|\bm{q}||z-z'|}-e^{-|\bm{q}|(z+z')}\right]-\right.\nonumber\\
&&\left.~~~\frac{q_{\alpha}q_{\beta}}{|\bm{q}|}\left[(|z-z'|e^{-|\bm{q}||z-z'|}-(z+z')e^{-|\bm{q}|(z+z')}\right]-
2zz'q_{\alpha}q_{\beta}e^{-|\bm{q}|(z+z')}\right\},\label{BlakeFTalphabeta}\nonumber \\
G_{\alpha z}({\bm R};{\bm R}')&=&\frac{1}{16\pi^{2}\eta}\int d^{2}q\frac{1}{|\bm{q}|}e^{i\bm{q}\cdot(\bm{R}_{p}-\bm{R}_{p}')}iq_{\alpha}\left[-(z-z')e^{-|\bm{q}||z-z'|}+(z-z')e^{-|\bm{q}|(z+z')}+2zz'|\bm{q}|e^{-|\bm{q}|(z+z')}
\right],\label{BlakeFTalphaz}\nonumber \\
G_{z\alpha}({\bm R};{\bm R}')&=&\frac{1}{16\pi^{2}\eta}\int d^{2}q\frac{1}{|\bm{q}|}e^{i\bm{q}\cdot(\bm{R}_{p}-\bm{R}_{p}')}iq_{\alpha}\left[-(z-z')e^{-|\bm{q}||z-z'|}+(z-z')e^{-|\bm{q}|(z+z')}-2zz'|\bm{q}|e^{-|\bm{q}|(z+z')}
\right],\label{BlakeFTzalpha}\nonumber \\
G_{zz}({\bm R};{\bm R}')&=&\frac{1}{16\pi^{2}\eta}\int d^{2}q\frac{1}{|\bm{q}|}e^{i\bm{q}\cdot(\bm{R}_{p}-\bm{R}_{p}')}\left[e^{-|\bm{q}|(z-z')}-e^{-|\bm{q}|(z+z')}+|\bm{q}||z-z'|e^{-|\bm{q}||z-z'|}\right.\nonumber\\
&&~~~-\left.|\bm{q}|(z+z')e^{-|\bm{q}|(z+z')}-2zz'|\bm{q}|^{2}e^{-|\bm{q}|(z+z')} \right], \label{eq:BlakeFT}
\end{eqnarray}
where $\alpha,\beta=x,y$.

\section{Writing the dynamical equation in the new coordinate}
A new coordinate is chosen that makes the intrinsic velocity constant: $d\bar{\phi}/d\phi=\Omega(\phi)/\Omega_0$. To linear order in harmonic amplitudes, the relation between old and new coordinates is
\begin{equation}
\phi(\bar{\phi})\simeq\bar{\phi}+\sum_{n=1}\frac{1}{n}\left[\left(A_{n}-C_n\right)\sin n\bar{\phi}-\left(B_{n}-D_n\right) \cos n\bar{\phi}\right].
\label{relationbetweenphase}
\end{equation}
Expanding out the harmonic terms to first order gives
\begin{subequations}
\begin{eqnarray}
   \cos{\Big(\bar{\phi}+\sum{H_n}\Big)}=\cos{\bar{\phi}}-\sin{\bar{\phi}}\sum{H_n}, \\
   \sin{\Big(\bar{\phi}+\sum{H_n}\Big)}=\sin{\bar{\phi}}+\cos{\bar{\phi}}\sum{H_n},
\end{eqnarray}
\end{subequations}
where we denote
\begin{eqnarray}
   H_n=\frac{1}{n}\left[\left(A_{n}-C_n\right)\sin n\bar{\phi}-\left(B_{n}-D_n\right) \cos n\bar{\phi}  \right], \nonumber \\
   H_n'=\frac{1}{n}\left[\left(A_{n}-C_n\right)\sin n\bar{\phi}'-\left(B_{n}-D_n\right) \cos n\bar{\phi}'  \right].
\end{eqnarray}
I also rewrite the equation in terms of phase differences and sum of phases:
\begin{equation}
\Delta=\bar{\phi}-\bar{\phi}', \quad \varphi=\bar{\phi}+\bar{\phi}', \quad \bar{\phi}=\frac{\varphi+\Delta}{2}, \quad \bar{\phi}'=\frac{\varphi-\Delta}{2}.
\label{eq:sepvars}
\end{equation}
The tangent of the rotor trajectories can be written as
\begin{eqnarray}
   &t_\alpha(\phi)&=-\sin{\phi} f_\alpha+\cos{\phi}g_\alpha, \nonumber \\
   &t_3(\phi)&=\cos{\phi}f_3,
\end{eqnarray}
where
\begin{eqnarray}
   f_1=\cos{\theta}, f_2=\sin{\theta}, f_3=\cos{\chi}, \nonumber \\
   g_1=\sin{\chi}\sin{\theta}, g_2=-\sin{\chi}\cos{\theta}
\end{eqnarray}
The dynamical equation for $\bar{\phi}$ is
\begin{equation}
   \frac{d \bar{\phi}}{d t}=\frac{\Omega_0 a \zeta(\phi(\bar{\phi}))}{F(\phi(\bar{\phi}))}\frac{d \phi}{d t}.
   \label{eq:dynamicalnew}
\end{equation}
\stepcounter{equation}
Now, we insert the expressions for the Blake tensor in Eq. \ref{eq:BlakeFT} into Eq. \ref{eq:dynamicalnew} using the rotor positions from Eq. 1 in the main text and forcing and friction from Eq. 3 in the main text. Next, expand the Blake tensor expressions to linear order in $a$, retain terms that are linear in the harmonic amplitudes and neglect higher order terms. This gives the following terms in the dynamical equation for $\bar{\phi}$:
\begin{align*}
   \frac{d\bar{\phi}}{dt}=\frac{F_0}{\zeta_0 a}+\frac{F_0}{a}\sum_{{\bm r}'}\left(1+\sum_n(C_n-A_n)\cos{\left(\frac{n}{2}(\varphi+\Delta)\right)}+(D_n-B_n)\sin{\left(\frac{n}{2}(\varphi+\Delta)\right)}\right)\Bigg(1+\sum_n A_n\cos{\left(\frac{n}{2}(\varphi-\Delta)\right)}\\
   +B_n\sin{\left(\frac{n}{2}(\varphi-\Delta)\right)} \Bigg)\int\frac{d^2{\bm q}}{16\pi^2\eta}\frac{1}{|{\bm q}|}e^{i{\bm q}\cdot({\bm r}-{\bm r}')}\Big(1+ia\big[q_\gamma f_\gamma(-2\sin{\frac{\varphi}{2}}\sin{\frac{\Delta}{2}}-\sum_m H_m\sin{\frac{\varphi+\Delta}{2}}+\sum_m H_m'\sin{\frac{\varphi-\Delta}{2}})\\
   +q\gamma g_\gamma(2\cos{\frac{\varphi}{2}}\sin{\frac{\Delta}{2}}+\sum_m H_m\cos{\frac{\varphi+\Delta}{2}}-\sum_m H_m'\cos{\frac{\varphi-\Delta}{2}})  \big]\Big) \Bigg\{\Big[\big(\sin{\frac{\varphi+\Delta}{2}}+\sum_m H_m\cos{\frac{\varphi+\Delta}{2}}\big)\big(\sin{\frac{\varphi-\Delta}{2}}\\
   +\sum_m H_m'\cos{\frac{\varphi-\Delta}{2}}\big)f_\alpha f_\beta-\big(\sin{\frac{\varphi+\Delta}{2}}+\sum_m H_m\cos{\frac{\varphi+\Delta}{2}}\big)\big(\cos{\frac{\varphi-\Delta}{2}}-\sum_m H_m'\sin{\frac{\varphi-\Delta}{2}}\big)f_\alpha g_\beta+\big(\cos{\frac{\varphi+\Delta}{2}}\\
   -\sum_m H_m\sin{\frac{\varphi+\Delta}{2}}\big)\big(\sin{\frac{\varphi-\Delta}{2}}+\sum_m H_m'\cos{\frac{\varphi-\Delta}{2}}\big)g_\alpha f_\beta+\big(\cos{\frac{\varphi+\Delta}{2}}-\sum_m H_m\sin{\frac{\varphi+\Delta}{2}}\big)\big(\cos{\frac{\varphi-\Delta}{2}}\\
   -\sum_m H_m'\sin{\frac{\varphi-\Delta}{2}}\big)g_\alpha g_\beta   \Big]\Bigg[\left(2\delta_{\alpha \beta}-\frac{q_\alpha q_\beta}{|{\bm q}|^2} \right)\Big(1-|{\bm q}|af_3\Big|2\cos{\frac{\varphi}{2}}\sin{\frac{\Delta}{2}}+\sum_m H_m\cos{\frac{\varphi+\Delta}{2}}-\sum_m H_m'\cos{\frac{\varphi-\Delta}{2}}\Big|\\
   -e^{-2|{\bm q}|h}\big(1-|{\bm q}|af_3(2\sin{\frac{\varphi}{2}}\cos{\frac{\Delta}{2}}+\sum_m H_m\cos{\frac{\varphi+\Delta}{2}}+\sum_m H_m'\cos{\frac{\varphi-\Delta}{2}})\big)\Big)-\frac{q_\alpha q_\beta}{|{\bm q}|}\Big(a f_3\Big|2\cos{\frac{\varphi}{2}}\sin{\frac{\Delta}{2}}\\
   +\sum_m H_m\cos{\frac{\varphi+\Delta}{2}}-\sum_m H_m'\cos{\frac{\varphi-\Delta}{2}}\Big|-\big(2h+af_3(2\sin{\frac{\varphi}{2}}\cos{\frac{\Delta}{2}}+\sum_m H_m\cos{\frac{\varphi+\Delta}{2}}+\sum_m H_m'\cos{\frac{\varphi-\Delta}{2}})  \big)\\
   e^{-2h|{\bm q}|}\big(1-|{\bm q}|af_3(2\sin{\frac{\varphi}{2}}\cos{\frac{\Delta}{2}}+\sum_m H_m\cos{\frac{\varphi+\Delta}{2}}+\sum_m H_m'\cos{\frac{\varphi-\Delta}{2}})\big) \Big)-2h\big(h+af_3(2\sin{\frac{\varphi}{2}}\cos{\frac{\Delta}{2}}\\
   +\sum_m H_m\cos{\frac{\varphi+\Delta}{2}}+\sum_m H_m'\cos{\frac{\varphi-\Delta}{2}}) \big)q_\alpha q_\beta e^{-2h|{\bm q}|}\big(1-|{\bm q}|af_3(2\sin{\frac{\varphi}{2}}\cos{\frac{\Delta}{2}}+\sum_m H_m\cos{\frac{\varphi+\Delta}{2}}\\
   +\sum_m H_m'\cos{\frac{\varphi-\Delta}{2}})\big)   \Bigg]+\Big[-(\sin{\frac{\varphi+\Delta}{2}}+\sum_m H_m\cos{\frac{\varphi+\Delta}{2}})(\cos{\frac{\varphi-\Delta}{2}}-\sum_m H_m'\sin{\frac{\varphi-\Delta}{2}})f_\alpha f_3\\
   +(\cos{\frac{\varphi+\Delta}{2}}-\sum_m H_m\sin{\frac{\varphi+\Delta}{2}})(\cos{\frac{\varphi-\Delta}{2}}-\sum_m H_m'\sin{\frac{\varphi-\Delta}{2}})g_\alpha f_3\Big]iq_\alpha\Bigg[af_3(2\cos{\frac{\varphi}{2}}\sin{\frac{\Delta}{2}}+\sum_m H_m\cos{\frac{\varphi+\Delta}{2}}\\
      -\sum_m H_m'\cos{\frac{\varphi-\Delta}{2}})(-1+e^{-2|{\bm q}|h})+2h\big(h+af_3(2\sin{\frac{\varphi}{2}}\cos{\frac{\Delta}{2}}+\sum_m H_m\cos{\frac{\varphi+\Delta}{2}}+\sum_m H_m'\cos{\frac{\varphi-\Delta}{2}})  \big)\\
   |{\bm q}|e^{-2h|{\bm q}|}\big(1-|{\bm{q}}|af_3(2\sin{\frac{\varphi}{2}}\cos{\frac{\Delta}{2}}+\sum_m H_m\cos{\frac{\varphi+\Delta}{2}}+\sum_m H_m'\cos{\frac{\varphi-\Delta}{2}})  \big)    \Bigg]+\Big[-(\cos{\frac{\varphi+\Delta}{2}}-\sum_m H_m\sin{\frac{\varphi+\Delta}{2}})\\
   (\sin{\frac{\varphi-\Delta}{2}}+\sum_m H_m\cos{\frac{\varphi-\Delta}{2}})f_3 f_\alpha+(\cos{\frac{\varphi+\Delta}{2}}-\sum_m H_m\sin{\frac{\varphi+\Delta}{2}})(\cos{\frac{\varphi-\Delta}{2}}-\sum_m H_m\sin{\frac{\varphi-\Delta}{2}}) f_3 g_\alpha \Big]\\
   iq_\alpha\Bigg[af_3(2\cos{\frac{\varphi}{2}}\sin{\frac{\Delta}{2}}+\sum_m H_m\cos{\frac{\varphi+\Delta}{2}}-\sum_m H_m'\cos{\frac{\varphi-\Delta}{2}})(-1+e^{-2h|{\bm q}|})-2h\big(h+af_3(2\sin{\frac{\varphi}{2}}\cos{\frac{\Delta}{2}}\\
   +\sum_m H_m\cos{\frac{\varphi+\Delta}{2}}+\sum_m H_m'\cos{\frac{\varphi-\Delta}{2}})  \big)|{\bm q}|e^{-2h|{\bm q}|}\big(1-|{\bm q}|af_3(2\sin{\frac{\varphi}{2}}\cos{\frac{\Delta}{2}}+\sum_m H_m\cos{\frac{\varphi+\Delta}{2}}+\sum_m H_m'\cos{\frac{\varphi-\Delta}{2}})  \big)  \Bigg]\\
   +(\cos{\frac{\varphi+\Delta}{2}}-\sum_m H_m\sin{\frac{\varphi+\Delta}{2}})(\cos{\frac{\varphi-\Delta}{2}}-\sum_m H_m'\sin{\frac{\varphi-\Delta}{2}}) f_3^2\Bigg[\big(1-|{\bm q}|af_3|2\cos{\frac{\varphi}{2}}\sin{\frac{\Delta}{2}}+\sum_m H_m\cos{\frac{\varphi+\Delta}{2}}\\
   -\sum_m H_m'\cos{\frac{\varphi-\Delta}{2}}| \big)-e^{-2h|{\bm q}|}\big(1-|{\bm q}|af_3(2\sin{\frac{\varphi}{2}}\cos{\frac{\Delta}{2}}+\sum_m H_m\cos{\frac{\varphi+\Delta}{2}}+\sum_m H_m'\cos{\frac{\varphi-\Delta}{2}}) \big)+|{\bm q}|af_3\big| 2\cos{\frac{\varphi}{2}}\sin{\frac{\Delta}{2}}\\
   +\sum_m H_m\cos{\frac{\varphi+\Delta}{2}}-\sum_m H_m'\cos{\frac{\varphi-\Delta}{2}}  \big|-|q|\big(2h+af_3(2\sin{\frac{\varphi}{2}}\cos{\frac{\Delta}{2}}+\sum_m H_m\cos{\frac{\varphi+\Delta}{2}}+\sum_m H_m'\cos{\frac{\varphi-\Delta}{2}}) \big)e^{-2h|{\bm q}|}\\
   \big(1-|{\bm q}|af_3(2\sin{\frac{\varphi}{2}}\cos{\frac{\Delta}{2}}+\sum_m H_m\cos{\frac{\varphi+\Delta}{2}}+\sum_m H_m'\cos{\frac{\varphi-\Delta}{2}})\big)-2h\big(h+af_3(2\sin{\frac{\varphi}{2}}\cos{\frac{\Delta}{2}}+\sum_m H_m\cos{\frac{\varphi+\Delta}{2}}\\
   +\sum_m H_m'\cos{\frac{\varphi-\Delta}{2}}) \big)|{\bm q}|^2e^{-2h|{\bm q}|}\big(1-|{\bm q}|af_3(2\sin{\frac{\varphi}{2}}\cos{\frac{\Delta}{2}}+\sum_m H_m\cos{\frac{\varphi+\Delta}{2}}+\sum_m H_m'\cos{\frac{\varphi-\Delta}{2}})\big)     \Bigg]   \Bigg\} 
   \label{eq:newcoord} 
   \tag{\theequation}
\end{align*}
\stepcounter{equation}

\section{Terms in the coarse-grained dynamical equation}

The phase difference between any pair of rotors, $\Delta=\bar{\phi}-\bar{\phi}'$, evolves on a much slower timescale than the average phase between any pair of rotors, $\varphi=(\bar{\phi}+\bar{\phi}')/2$, so we time average over fast variables to give the dynamical equation only in terms of phase differences, $\Delta$. Worked examples of how to carry out the time-averaging calculation are shown in the SI of Ref. \cite{Meng:2021}.
Only the first and second harmonics survive the time-averaging process. Contributions to the dynamical equation from the first and second harmonic are given in the following two subsections.

\subsection{First harmonic terms that survive time-averaging over the fast variable}
\begin{align*} 
\dot{\bar{\phi}}_{<n=1~{\rm terms}>}=\frac{F_0}{a}\sum_{{\bm r}'}\int\frac{d^2{\bm q}}{16\pi^2\eta}\frac{1}{|{\bm q}|}e^{i{\bm q}\cdot({\bm r}-{\bm r}')}\Bigg(\Bigg[\Big(\frac{ia}{2}\Big) \Big\{\big[(-q_\gamma f_\gamma(-C_1+2A_1)\\
   +q\gamma g_\gamma(D_1-2B_1))\Big(\Big[\frac{1}{2} (f_\alpha f_\beta+g_\alpha g_\beta) \Big]\Big)\big]\\
    \Bigg[\left(2\delta_{\alpha \beta}-\frac{q_\alpha q_\beta}{|{\bm q}|^2} \right)\Big(1-e^{-2|{\bm q}|h}\Big)+\frac{q_\alpha q_\beta}{|{\bm q}|}\Big(2h e^{-2h|{\bm q}|}\Big)-2h^2q_\alpha q_\beta e^{-2h|{\bm q}|}  \Bigg]\\
   +\big[-q_\gamma f_\gamma(-C_1+2A_1)+q\gamma g_\gamma(D_1-2B_1)  \big]\frac{1}{2}  f_3^2\Bigg[1-e^{-2h|{\bm q}|}-|{\bm q}|\big(2h \big)e^{-2h|{\bm q}|}-2h^2|{\bm q}|^2e^{-2h|{\bm q}|}     \Bigg]   \Big\}\\
   -{\rm i}a\big[q_\gamma f_\gamma  \big]\Big\{(B_1-D_1)  f_\alpha g_\beta \Bigg[\left(2\delta_{\alpha \beta}-\frac{q_\alpha q_\beta}{|{\bm q}|^2} \right)\Big(1-e^{-2|{\bm q}|h}\Big)+\frac{q_\alpha q_\beta}{|{\bm q}|}\Big(2h e^{-2h|{\bm q}|}\Big)-2h^2 q_\alpha q_\beta e^{-2h|{\bm q}|}  \Bigg]\\
   +(B_1-D_1) f_\alpha f_3 {\rm i}q_\alpha\Bigg[2h^2|{\bm q}|e^{-2h|{\bm q}|}    \Bigg] \Big\}\\
   -{\rm i}a\big[q\gamma g_\gamma  \big]\Big\{(A_1-C_1)f_\alpha g_\beta    \Bigg[\left(2\delta_{\alpha \beta}-\frac{q_\alpha q_\beta}{|{\bm q}|^2} \right)\Big(1-e^{-2|{\bm q}|h}\Big)+\frac{q_\alpha q_\beta}{|{\bm q}|}\Big(2h e^{-2h|{\bm q}|}\Big)-2h^2 q_\alpha q_\beta e^{-2h|{\bm q}|}  \Bigg]\\
   +(A_1-C_1)f_\alpha f_3 {\rm i}q_\alpha\Bigg[2h^2|{\bm q}|e^{-2h|{\bm q}|}    \Bigg]   \Big\}\\
   +\Big[\frac{1}{2} (f_\alpha f_\beta+g_\alpha g_\beta)D_1 \Big] \Bigg[\delta_{\alpha \beta} e^{-2|{\bm q}|h}|{\bm q}|af_3-ha f_3\big(-h|{\bm q}|+2 \big)q_\alpha q_\beta e^{-2h|{\bm q}|}   \Bigg]\\
   + g_\alpha f_3\frac{{\rm i}q_\alpha}{2}\Bigg[af_3(D_1-2B_1)(-1+e^{-2|{\bm q}|h}) \Bigg] +\frac{1}{2} f_3^2 D_1 \Big[h^2 af_3|{\bm q}|^3{\rm e}^{-2h|{\bm q}|}     \Big] \Bigg]\cos{\Delta} \\ 
+\Bigg[ \Big(\frac{ia}{2}\Big) \Big\{\big[(-q_\gamma f_\gamma(-C_1+2A_1)+q\gamma g_\gamma(D_1-2B_1))(- f_\alpha g_\beta  )  \big]\\
    \Bigg[\left(2\delta_{\alpha \beta}-\frac{q_\alpha q_\beta}{|{\bm q}|^2} \right)\Big(1-e^{-2|{\bm q}|h}\Big)+\frac{q_\alpha q_\beta}{|{\bm q}|}\Big(2h e^{-2h|{\bm q}|}\Big)-2h^2q_\alpha q_\beta e^{-2h|{\bm q}|}  \Bigg]\\
   +\big[-q_\gamma f_\gamma(-C_1+2A_1)+q\gamma g_\gamma(D_1-2B_1)  \big]\Big[-  f_\alpha f_3\Big]iq_\alpha\Bigg[2h^2|{\bm q}|e^{-2h|{\bm q}|}    \Bigg]  \Big\}\\
   +\frac{ia}{2}\big[-q_\gamma f_\gamma  \big] \Big\{(B_1-D_1)\Big[  (f_\alpha f_\beta+g_\alpha g_\beta)  \Big] \\
   \Bigg[\left(2\delta_{\alpha \beta}-\frac{q_\alpha q_\beta}{|{\bm q}|^2} \right)\Big(1-e^{-2|{\bm q}|h}\Big)+\frac{q_\alpha q_\beta}{|{\bm q}|}\Big(2h e^{-2h|{\bm q}|}\Big)-2h^2 q_\alpha q_\beta e^{-2h|{\bm q}|}  \Bigg]\\
+(B_1-D_1) f_3^2\Bigg[1-e^{-2h|{\bm q}|}-|{\bm q}|\big(2h \big)e^{-2h|{\bm q}|}-2h^2|{\bm q}|^2e^{-2h|{\bm q}|}     \Bigg]   \Big\}\\
   +\frac{ia}{2}\big[q\gamma g_\gamma  \big]\Big\{(A_1-C_1)\Big[\big(\big(-(f_\alpha f_\beta+g_\alpha g_\beta)  \Big] \\
   \Bigg[\left(2\delta_{\alpha \beta}-\frac{q_\alpha q_\beta}{|{\bm q}|^2} \right)\Big(1-e^{-2|{\bm q}|h}\Big)+\frac{q_\alpha q_\beta}{|{\bm q}|}\Big(2h e^{-2h|{\bm q}|}\Big)-2h^2 q_\alpha q_\beta e^{-2h|{\bm q}|}  \Bigg]\\
   -(A_1-C_1) f_3^2\Bigg[1-e^{-2h|{\bm q}|}-|{\bm q}|\big(2h \big)e^{-2h|{\bm q}|}-2h^2|{\bm q}|^2e^{-2h|{\bm q}|}     \Bigg]   \Big\}\\
   - f_\alpha g_\beta D_1 \Bigg[\delta_{\alpha \beta} e^{-2|{\bm q}|h}|{\bm q}|af_3-ha f_3\big(-h|{\bm q}|+2 \big)q_\alpha q_\beta e^{-2h|{\bm q}|}   \Bigg]\\
   - f_\alpha f_3 {\rm i}q_\alpha\Big[haf_3 D_1 \big(-|{\bm q}|h+1  \big)|{\bm q}|e^{-2h|{\bm q}|}    \Big] \Bigg]\sin{\Delta} \\ 
+\Bigg[\Big(\frac{ia}{2}\Big) \Big\{\big[(q_\gamma f_\gamma(-C_1+2A_1)-q\gamma g_\gamma(D_1-2B_1))\frac{1}{4} (f_\alpha f_\beta+g_\alpha g_\beta)  \\
   +(-q_\gamma f_\gamma D_1+q\gamma g_\gamma C_1)\frac{1}{2}  f_\alpha g_\beta    \big]\\
    \Bigg[\left(2\delta_{\alpha \beta}-\frac{q_\alpha q_\beta}{|{\bm q}|^2} \right)\Big(1-e^{-2|{\bm q}|h}\Big)+\frac{q_\alpha q_\beta}{|{\bm q}|}\Big(2h e^{-2h|{\bm q}|}\Big)-2h^2q_\alpha q_\beta e^{-2h|{\bm q}|}  \Bigg]\\
   +\big[\frac{1}{2}q_\gamma f_\gamma D_1+\frac{1}{2}q\gamma g_\gamma C_1  \big]\Big[-  f_\alpha f_3\Big]iq_\alpha\Bigg[2h^2|{\bm q}|e^{-2h|{\bm q}|}    \Bigg]+\big[\frac{1}{2}q_\gamma f_\gamma(-C_1+2A_1)\\
   -\frac{1}{2}q\gamma g_\gamma(D_1-2B_1)  \big]\frac{1}{2}  f_3^2\Bigg[1-e^{-2h|{\bm q}|}-|{\bm q}|\big(2h \big)e^{-2h|{\bm q}|}-2h^2|{\bm q}|^2e^{-2h|{\bm q}|}     \Bigg]   \Big\}\\
   +\frac{ia}{2}\big[q_\gamma f_\gamma  \big]\Big\{(B_1-D_1)  f_\alpha g_\beta     \Bigg[\left(2\delta_{\alpha \beta}-\frac{q_\alpha q_\beta}{|{\bm q}|^2} \right)\Big(1-e^{-2|{\bm q}|h}\Big)+\frac{q_\alpha q_\beta}{|{\bm q}|}\Big(2h e^{-2h|{\bm q}|}\Big)-2h^2 q_\alpha q_\beta e^{-2h|{\bm q}|}  \Bigg]\\
   -(B_1-D_1) f_\alpha f_3 {\rm i}q_\alpha\Bigg[2h^2|{\bm q}|e^{-2h|{\bm q}|}    \Bigg] \Big\}\\
   +\frac{ia}{2}\big[q\gamma g_\gamma  \big]\Big\{(A_1-C_1) f_\alpha g_\beta   \Bigg[\left(2\delta_{\alpha \beta}-\frac{q_\alpha q_\beta}{|{\bm q}|^2} \right)\Big(1-e^{-2|{\bm q}|h}\Big)+\frac{q_\alpha q_\beta}{|{\bm q}|}\Big(2h e^{-2h|{\bm q}|}\Big)-2h^2 q_\alpha q_\beta e^{-2h|{\bm q}|}  \Bigg]\\
   +(A_1-C_1) f_\alpha f_3 {\rm i}q_\alpha\Bigg[2h^2|{\bm q}|e^{-2h|{\bm q}|}    \Bigg] \Big\}\\
   +\Big[\frac{1}{4} (f_\alpha f_\beta+g_\alpha g_\beta)D_1- \frac{1}{2}f_\alpha g_\beta (C_1-2A_1)  \Big]\Bigg[\delta_{\alpha \beta} e^{-2|{\bm q}|h}|{\bm q}|af_3-ha f_3\big(-h|{\bm q}|+2 \big)q_\alpha q_\beta e^{-2h|{\bm q}|}   \Bigg]\\
   - f_\alpha f_3 {\rm i}q_\alpha\Bigg[haf_3\Big((C_1-2A_1)\frac{1}{2}\Big)\big(-|{\bm q}|h+1  \big)|{\bm q}|e^{-2h|{\bm q}|}    \Bigg] \\
   - g_\alpha f_3\frac{{\rm i}q_\alpha}{4}\Bigg[af_3(D_1-2B_1)(-1+e^{-2|{\bm q}|h}) \Bigg] +\frac{1}{4} f_3^2 D_1\Big[h^2 af_3|{\bm q}|^3{\rm e}^{-2h|{\bm q}|}     \Big] \Bigg]\cos{2\Delta} \\
+ \Bigg[ 
\Big(\frac{ia}{2}\Big) \Big\{\big[(-q_\gamma f_\gamma(-C_1+2A_1)+q\gamma g_\gamma(D_1-2B_1))\big( \frac{1}{2} f_\alpha g_\beta   \big)\\
   +(-q_\gamma f_\gamma D_1+q\gamma g_\gamma C_1)\Big[\frac{1}{4}(f_\alpha f_\beta+g_\alpha g_\beta)    \Big]\big]\\
    \Bigg[\left(2\delta_{\alpha \beta}-\frac{q_\alpha q_\beta}{|{\bm q}|^2} \right)\Big(1-e^{-2|{\bm q}|h}\Big)+\frac{q_\alpha q_\beta}{|{\bm q}|}\Big(2h e^{-2h|{\bm q}|}\Big)-2h^2q_\alpha q_\beta e^{-2h|{\bm q}|}  \Bigg]\\
   +\big[-q_\gamma f_\gamma(-C_1+2A_1)+q\gamma g_\gamma(D_1-2B_1)  \big]\Big[ f_\alpha f_3\Big]iq_\alpha\Bigg[h^2|{\bm q}|e^{-2h|{\bm q}|}    \Bigg]\\
   +\big[-q_\gamma f_\gamma D_1+q\gamma g_\gamma C_1  \big]\frac{1}{4}  f_3^2\Bigg[1-e^{-2h|{\bm q}|}-|{\bm q}|\big(2h \big)e^{-2h|{\bm q}|}-2h^2|{\bm q}|^2e^{-2h|{\bm q}|}     \Bigg]   \Big\}\\
   +\frac{ia}{4}\big[q_\gamma f_\gamma  \big]\Big\{(B_1-D_1) (f_\alpha f_\beta+g_\alpha g_\beta)  \\
   \Bigg[\left(2\delta_{\alpha \beta}-\frac{q_\alpha q_\beta}{|{\bm q}|^2} \right)\Big(1-e^{-2|{\bm q}|h}\Big)+\frac{q_\alpha q_\beta}{|{\bm q}|}\Big(2h e^{-2h|{\bm q}|}\Big)-2h^2 q_\alpha q_\beta e^{-2h|{\bm q}|}  \Bigg]\\
   -\frac{1}{2}(B_1-D_1) f_3^2\Big[1-e^{-2h|{\bm q}|}-|{\bm q}|\big(2h \big)e^{-2h|{\bm q}|}-2h^2|{\bm q}|^2e^{-2h|{\bm q}|}     \Big]   \Big\}\\
   +\frac{ia}{4}\big[q\gamma g_\gamma  \big]\Big\{(A_1-C_1)(f_\alpha f_\beta+g_\alpha g_\beta)   \\
   \Bigg[\left(2\delta_{\alpha \beta}-\frac{q_\alpha q_\beta}{|{\bm q}|^2} \right)\Big(1-e^{-2|{\bm q}|h}\Big)+\frac{q_\alpha q_\beta}{|{\bm q}|}\Big(2h e^{-2h|{\bm q}|}\Big)-2h^2 q_\alpha q_\beta e^{-2h|{\bm q}|}  \Bigg]\\
   +\frac{1}{2}(A_1-C_1) f_3^2\Bigg[1-e^{-2h|{\bm q}|}-|{\bm q}|\big(2h \big)e^{-2h|{\bm q}|}-2h^2|{\bm q}|^2e^{-2h|{\bm q}|}     \Bigg]   \Big\}\\
   -\frac{1}{2}\Big(\frac{1}{2} (f_\alpha f_\beta+g_\alpha g_\beta)(C_1-2A_1)+ f_\alpha g_\beta D_1   \Big)\Big[\delta_{\alpha \beta} e^{-2|{\bm q}|h}|{\bm q}|af_3-ha f_3\big(-h|{\bm q}|+2 \big)q_\alpha q_\beta e^{-2h|{\bm q}|}   \Big]\\
   - \frac{1}{2}f_\alpha f_3 {\rm i}q_\alpha [haf_3 D_1 \big(-|{\bm q}|h+1  \big)|{\bm q}|e^{-2h|{\bm q}|}   ] + g_\alpha f_3\frac{{\rm i}q_\alpha}{4}[af_3 C_1(-1+e^{-2|{\bm q}|h}) ] \\
   -\frac{1}{2} f_3^2(C_1-2A_1)\Big[h^2 af_3|{\bm q}|^3{\rm e}^{-2h|{\bm q}|}     \Big]  
  \Bigg]\sin{2\Delta}\\ 
+\Bigg[\Big(\frac{ia}{2}\Big) \Big\{\big[(-q_\gamma f_\gamma(-C_1+2A_1)+q\gamma g_\gamma(D_1-2B_1))\Big(-\frac{1}{4}(f_\alpha f_\beta+g_\alpha g_\beta)  \Big)\\
   +(-q_\gamma f_\gamma D_1+q\gamma g_\gamma C_1)\Big[-\frac{1}{2}  f_\alpha g_\beta    \Big]\big]\\
    \Bigg[\left(2\delta_{\alpha \beta}-\frac{q_\alpha q_\beta}{|{\bm q}|^2} \right)\Big(1-e^{-2|{\bm q}|h}\Big)+\frac{q_\alpha q_\beta}{|{\bm q}|}\Big(2h e^{-2h|{\bm q}|}\Big)-2h^2q_\alpha q_\beta e^{-2h|{\bm q}|}  \Bigg]\\
   +\big[-\frac{1}{2}q_\gamma f_\gamma D_1+q\gamma g_\gamma C_1\frac{1}{2}  \big]\Big[-  f_\alpha f_3\Big]iq_\alpha\Bigg[2h^2|{\bm q}|e^{-2h|{\bm q}|}    \Bigg]\\
   +\big[\frac{1}{2} q_\gamma f_\gamma(-C_1+2A_1)-\frac{1}{2}q\gamma g_\gamma(D_1-2B_1)  \big]\frac{1}{2}  f_3^2\Bigg[1-e^{-2h|{\bm q}|}-|{\bm q}|\big(2h \big)e^{-2h|{\bm q}|}-2h^2|{\bm q}|^2e^{-2h|{\bm q}|}     \Bigg]   \Big\}\\
   +\frac{ia}{2}\big[-q_\gamma f_\gamma  \big] \Big\{(B_1-D_1)[  -f_\alpha g_\beta  ] \\
   \Bigg[\left(2\delta_{\alpha \beta}-\frac{q_\alpha q_\beta}{|{\bm q}|^2} \right)\Big(1-e^{-2|{\bm q}|h}\Big)+\frac{q_\alpha q_\beta}{|{\bm q}|}\Big(2h e^{-2h|{\bm q}|}\Big)-2h^2 q_\alpha q_\beta e^{-2h|{\bm q}|}  \Bigg]\\
   +(B_1-D_1)\Big[- f_\alpha f_3\Big]iq_\alpha\Bigg[2h^2|{\bm q}|e^{-2h|{\bm q}|}    \Bigg] \Big\}+\frac{ia}{2}q_\gamma g_\gamma \Big\{(A_1-C_1)\Big[f_\alpha g_\beta  \Big] \\
   \Bigg[\left(2\delta_{\alpha \beta}-\frac{q_\alpha q_\beta}{|{\bm q}|^2} \right)\Big(1-e^{-2|{\bm q}|h}\Big)+\frac{q_\alpha q_\beta}{|{\bm q}|}\Big(2h e^{-2h|{\bm q}|}\Big)-2h^2 q_\alpha q_\beta e^{-2h|{\bm q}|}  \Bigg]\\
   +(A_1-C_1)\Big[ f_\alpha f_3\Big]iq_\alpha\Bigg[2h^2|{\bm q}|e^{-2h|{\bm q}|}    \Bigg]  \Big\}+\Big[\frac{1}{4} (f_\alpha f_\beta+g_\alpha g_\beta)D_1+\frac{1}{2} f_\alpha g_\beta (C_1-2A_1)  \Big]\\
   \Bigg[\delta_{\alpha \beta} e^{-2|{\bm q}|h}|{\bm q}|af_3-ha f_3\big(-h|{\bm q}|+2 \big)q_\alpha q_\beta e^{-2h|{\bm q}|}   \Bigg]\\
   - f_\alpha f_3 \frac{{\rm i}q_\alpha}{2}\Bigg[ h a f_3(-C_1+2A_1)\big(-|{\bm q}|h+1  \big)|{\bm q}|e^{-2h|{\bm q}|}    \Bigg] \\
   + g_\alpha f_3\frac{{\rm i}q_\alpha}{4}\Bigg[af_3(-D_1+2B_1)(-1+e^{-2|{\bm q}|h}) \Bigg] +\frac{1}{4} f_3^2 D_1 \Big[h^2 af_3|{\bm q}|^3{\rm e}^{-2h|{\bm q}|}     \Big] \Bigg]  \Bigg).  \tag{\theequation}
\label{eq:all_n1_time_averaged_simplified}
\end{align*}
This can be written as
\begin{equation}
   \dot{\bar{\phi}}_{<n=1~{\rm terms}>}=\frac{\Omega_0 b}{4\pi}\sum_{{\bm r}'}\int d^2{\bm q}{\rm e}^{i{\bm q}\cdot({\bm r}-{\bm r}')}\frac{a}{\ell}\Bigg(\mathcal{M}_1({\bm q})\cos{\Delta}+\mathcal{S}_1({\bm q})\sin{\Delta}+K_1({\bm q})\cos{2\Delta}+J_1({\bm q})\sin{2\Delta}+P_1(\bm q) \Bigg) 
\end{equation}
\stepcounter{equation}

\subsection{Second harmonic terms that survive time-averaging over the fast variable}

\begin{align*}
   \dot{\bar{\phi}}_{<(n=2~{\rm terms})>}=\frac{F_0}{16\pi^2\eta a}\sum_{{\bm r}'}
\int\frac{d^2{\bm q}}{|{\bm q}|}e^{i{\bm q}\cdot({\bm r}-{\bm r}')}\Bigg\{\Bigg( \Big[\frac{1}{2}f_\alpha f_\beta+\frac{1}{2}g_\alpha g_\beta  \Big] \\
   \Big[\left(2\delta_{\alpha \beta}-\frac{q_\alpha q_\beta}{|{\bm q}|^2} \right)\Big(1-e^{-2|{\bm q}|h}\Big)+\frac{q_\alpha q_\beta}{|{\bm q}|}\big(2h {\rm e}^{-2h|{\bm q}|} \big)-2h^2 q_\alpha q_\beta {\rm e}^{-2h|{\bm q}|}   \Big]\\
   +\frac{1}{2} f_3^2\Big[1 -e^{-2h|{\bm q}|}-|{\bm q}|\big(2h \big){\rm e}^{-2h|{\bm q}|}-2h^2|{\bm q}|^2 {\rm e}^{-2h|{\bm q}|}   \Big]  \\
   +\Big[(-\frac{1}{4}C_2+ \frac{1}{2} (A_2-C_2))(f_\alpha f_\beta-g_\alpha g_\beta )   \Big] \\
   \Big[\Big(2\delta_{\alpha \beta}-\frac{q_\alpha q_\beta}{|{\bm q}|^2} \Big)\Big(1-e^{-2|{\bm q}|h}\Big)+\frac{q_\alpha q_\beta}{|{\bm q}|}\big(2h {\rm e}^{-2h|{\bm q}|} \big)-2h^2 q_\alpha q_\beta {\rm e}^{-2h|{\bm q}|}   \Big]\\
   +(\frac{1}{4}C_2-\frac{1}{2}(A_2-C_2)) f_3^2\Big[1 -e^{-2h|{\bm q}|}-|{\bm q}|\big(2h \big){\rm e}^{-2h|{\bm q}|}-2h^2|{\bm q}|^2 {\rm e}^{-2h|{\bm q}|}   \Big] \Bigg)\cos{\Delta} \\ 
   +\Bigg(\Big[-\frac{1}{2}f_\alpha g_\beta-\frac{1}{2}g_\alpha f_\beta\Big]\\
   \Bigg[\left(2\delta_{\alpha \beta}-\frac{q_\alpha q_\beta}{|{\bm q}|^2} \right)\Big(1-e^{-2|{\bm q}|h}\Big)+\frac{q_\alpha q_\beta}{|{\bm q}|}\big(2h {\rm e}^{-2h|{\bm q}|} \big)-2h^2 q_\alpha q_\beta {\rm e}^{-2h|{\bm q}|}   \Bigg]\\
   -f_\alpha f_3 {\rm i}q_\alpha\Bigg[2h^2|{\bm q}|e^{-2h|{\bm q}|}    \Bigg]+\Big[-\frac{1}{4}(D_2-2B_2)f_\alpha f_\beta +\frac{1}{4}(D_2-2B_2)g_\alpha g_\beta   \Big] \\
    \Bigg[\left(2\delta_{\alpha \beta}-\frac{q_\alpha q_\beta}{|{\bm q}|^2} \right)\Big(1-e^{-2|{\bm q}|h}\Big)+\frac{q_\alpha q_\beta}{|{\bm q}|}\big(2h {\rm e}^{-2h|{\bm q}|} \big)-2h^2 q_\alpha q_\beta {\rm e}^{-2h|{\bm q}|}   \Bigg]\\
   +\frac{1}{4}(D_2-2B_2) f_3^2\Bigg[1 -e^{-2h|{\bm q}|}-|{\bm q}|\big(2h \big){\rm e}^{-2h|{\bm q}|}-2h^2|{\bm q}|^2 {\rm e}^{-2h|{\bm q}|}   \Bigg] \Bigg)\sin{\Delta} \Bigg\}\\
   =\frac{\Omega_0 b}{4 \pi}\sum_{r'}\int d^2{\bm q}{\rm e}^{{\rm i}\bm q \cdot (\bm{r-r'})}\big[\mathcal{M}_2(\bm q)\cos{\Delta}+\mathcal{S}_2(\bm q)\sin{\Delta}  \big], \tag{\theequation}
   \label{eq:n2contr_i_time_averaged_both_terms}
\end{align*}

\section{Global travelling wave}
To look for global metachronal coordination, we look at wave solutions of the form
\begin{equation}
   \bar{\phi}=\omega t-{\bm k}\cdot {\bm r}+\delta\bar{\phi}_{\bm k}({\bm r},t),
   \label{eq:waveansatz}
\end{equation}
where $\delta\bar{\phi}_{\bm k}({\bm r},t)$ is a perturbation to the travelling wave with wavevector ${\bm k}$. We then have
\begin{equation}
   \Delta={\bm k}\cdot({\bm r'-\bm r})+\delta \bar{\phi}_k({\bm r},t)-\delta \bar{\phi}_k({\bm r'},t).
\end{equation}
We can approximate the harmonic terms as:
\begin{subequations}
\begin{eqnarray}
   \cos{\Delta}=\frac{{\rm e}^{i{\bm k}\cdot({\bm r'-\bm r})}+{\rm e}^{-i{\bm k}\cdot({\bm r'-\bm r})}}{2}-\frac{{\rm e}^{i{\bm k}\cdot({\bm r'-\bm r})}-{\rm e}^{-i{\bm k}\cdot({\bm r'-\bm r})}}{2{\rm i}}(\delta\bar{\phi}-\delta\bar{\phi}')  \\
   \sin{\Delta}=\frac{{\rm e}^{i{\bm k}\cdot({\bm r'-\bm r})}-{\rm e}^{-i{\bm k}\cdot({\bm r'-\bm r})}}{2{\rm i}}+\frac{{\rm e}^{i{\bm k}\cdot({\bm r'-\bm r})}+{\rm e}^{-i{\bm k}\cdot({\bm r'-\bm r})}}{2}(\delta\bar{\phi}-\delta\bar{\phi}') \\
   \cos{(2\Delta)}=\frac{{\rm e}^{2i{\bm k}\cdot({\bm r'-\bm r})}+{\rm e}^{-2i{\bm k}\cdot({\bm r'-\bm r})}}{2}-\frac{{\rm e}^{2i{\bm k}\cdot({\bm r'-\bm r})}-{\rm e}^{-2i{\bm k}\cdot({\bm r'-\bm r})}}{{\rm i}}(\delta\bar{\phi}-\delta\bar{\phi}')  \\
   \sin{(2\Delta)}=\frac{{\rm e}^{2i{\bm k}\cdot({\bm r'-\bm r})}-{\rm e}^{-2i{\bm k}\cdot({\bm r'-\bm r})}}{2{\rm i}}+{\rm e}^{2i{\bm k}\cdot({\bm r'-\bm r})}+{\rm e}^{-2i{\bm k}\cdot({\bm r'-\bm r})}(\delta\bar{\phi}-\delta\bar{\phi}') 
\end{eqnarray}
\end{subequations}
\stepcounter{equation}

\subsection{Dispersion relation}
We seek a dispersion relation by inserting Eq. \ref{eq:waveansatz} into Eq. 6 
from the main text, using the expressions given in Eqs. \ref{eq:all_n1_time_averaged_simplified}, \ref{eq:n2contr_i_time_averaged_both_terms}, and expanding to zeroth order in the perturbation $\delta\bar{\phi}_{\bm k}({\bm r},t)$. This gives us
\begin{align*}
   \omega=\Omega_0+\frac{\Omega_0 b}{4\pi}\sum_{{\bm r}'}\int {\rm d}^2{\bm q}{\rm e}^{{\rm i}{\bm q}\cdot({\bm r}-{\bm r}')}\Bigg[\frac{M({\bm q})}{2}\big({\rm e}^{-{\rm i}{\bm k}\cdot({\bm r}-{\bm r}')}+{\rm e}^{{\rm i}{\bm k}\cdot({\bm r}-{\bm r}')}\big)+\frac{S({\bm q})}{2{\rm i}}\big({\rm e}^{-{\rm i}{\bm k}\cdot({\bm r}-{\bm r}')}-{\rm e}^{{\rm i}{\bm k}\cdot({\bm r}-{\bm r}')}\big) \\
   +\frac{a}{\ell}\frac{K_1({\bm q})}{2}\big({\rm e}^{-2{\rm i}{\bm k}\cdot({\bm r}-{\bm r}')}+{\rm e}^{2{\rm i}{\bm k}\cdot({\bm r}-{\bm r}')}\big)+\frac{a}{\ell}\frac{J_1({\bm q})}{2{\rm i}}\big({\rm e}^{-2{\rm i}{\bm k}\cdot({\bm r}-{\bm r}')}-{\rm e}^{2{\rm i}{\bm k}\cdot({\bm r}-{\bm r}')}\big)+\frac{a}{\ell}P({\bm q})  \Bigg] \tag{\theequation}
   \label{eq:dispersion1}
\end{align*}
The sum is over an infinite two-dimensional lattice, so we can using the identity
\begin{equation}
   \sum_{{\bm r}'}{\rm e}^{{\rm i}{\bm q}\cdot({\bm r}-{\bm r}')}=\sum_{\bm G}\frac{4\pi}{\ell^2}\delta^2({\bm q}-{\bm G})
\end{equation}
\stepcounter{equation}
and write Eq. \ref{eq:dispersion1} as a sum over reciprocal lattice vectors, ${\bm G}$. (The reciprocal lattice ${\bm G}$ is the Fourier transform of the real space lattice, ${\bm r}$, with ${\bm r}\cdot {\bm G}=2\pi n, n \in \mathbb{Z}$.)
Performing the ${\bm q}$ integral using the Dirac delta functions gives

\begin{align*}
   \omega=\Omega_0+\frac{\Omega_0 b \pi}{\ell^2}\sum_{{\bm G}}\Bigg[\frac{M({\bm k}+{\bm G})+M(-{\bm k}+{\bm G})}{2}+\frac{S({\bm k}+{\bm G})-S(-{\bm k}+{\bm G})}{2{\rm i}}\ \\
   +\frac{a}{\ell}\frac{K_1(2{\bm k}+{\bm G})+K_1(-2{\bm k}+{\bm G})}{2}+\frac{a}{\ell}\frac{J_1(2{\bm k}+{\bm G})-J_1(-2{\bm k}+{\bm G})}{2{\rm i}}+\frac{a}{\ell}P_1({\bm G})  \Bigg] \tag{\theequation}
   \label{eq:dispersion2} 
\end{align*}
\stepcounter{equation}

\subsection{Linear stability analysis}
Expanding the dynamical equation with the wave ansatz Eq. \ref{eq:waveansatz} to first order in the perturbation gives the linear stability analysis. 
Expanding Eq. 6 from the main text 
to first order in the perturbation gives
\begin{align*}
   \partial_t \delta\bar{\phi}({\bm r},t)=\frac{\Omega_0 b}{4\pi}\sum_{{\bm r}'}\int {\rm d}^2{\bm q}{\rm e}^{{\rm i}{\bm q}\cdot({\bm r}-{\bm r}')}\Bigg[-\frac{M({\bm q})}{2{\rm i}}\big({\rm e}^{-{\rm i}{\bm k}\cdot({\bm r}-{\bm r}')}-{\rm e}^{{\rm i}{\bm k}\cdot({\bm r}-{\bm r}')}\big)+\frac{S({\bm q})}{2}\big({\rm e}^{-{\rm i}{\bm k}\cdot({\bm r}-{\bm r}')}+{\rm e}^{{\rm i}{\bm k}\cdot({\bm r}-{\bm r}')}\big) \\
   -\frac{a}{\ell}\frac{K_1({\bm q})}{{\rm i}}\big({\rm e}^{-2{\rm i}{\bm k}\cdot({\bm r}-{\bm r}')}-{\rm e}^{2{\rm i}{\bm k}\cdot({\bm r}-{\bm r}')}\big)+\frac{a}{\ell}J_1({\bm q})\big({\rm e}^{-2{\rm i}{\bm k}\cdot({\bm r}-{\bm r}')}+{\rm e}^{2{\rm i}{\bm k}\cdot({\bm r}-{\bm r}')}\big) \Bigg]\big(\delta\bar{\phi}({\bm r},t)-\delta\bar{\phi}({\bm r}',t) \big). \tag{\theequation}
\end{align*}
This has the form
\begin{equation}
   \partial_t \delta\bar{\phi}({\bm r},t)=\sum_{{\bm r}'}\gamma_{\bm k}({\bm r}-{\bm r}')(\delta\bar{\phi}({\bm r},t)-\delta\bar{\phi}({\bm r}',t) )
\end{equation}
Using the discrete Fourier transform of the perturbation $\delta\tilde{\phi}({\bm q}',t)=\sum_{\bm r'}{\rm e}^{{\rm i}{\bm q}'\cdot {\bm r}'}\delta\bar{\phi}({\bm r}',t)$ then gives us
\begin{equation}
   \partial_t \delta\tilde{\phi}({\bm q}',t)=\big( \tilde{\gamma}_{\bm k}(0)-\tilde{\gamma}_{\bm k}({\bm q}')\big) \delta\tilde{\phi}_{\bm k}({\bm q}',t)
\end{equation}
\stepcounter{equation}
which is the linear stability equation for the perturbation to a metachronal wave with wavevector ${\bm k}$ and perturbation wavevector ${\bm q}'$. For the metachronal wave with wavevector ${\bm k}$ to be globally linearly stable, we require $\tilde{\gamma}_{\bm k}(0)-\tilde{\gamma}_{\bm k}({\bm q}')<0$ for all perturbation wavevectors ${\bm q}'$.
The (discrete) Fourier transform of $\gamma_{\bm k}({\bm r})$ is $\tilde{\gamma}_{\bm k}({\bm q}')=\sum_{\bm r}{\rm e}^{{\rm i}{\bm q}'\cdot {\bm r}'}\gamma_{\bm k}(\bm r)$. This is given by
\begin{align*}
   \tilde{\gamma}_{\bm k}({\bm q}')=\frac{\Omega_0 b}{4\pi}\sum_{{\bm r}}{\rm e}^{{\rm i}{\bm q}'\cdot{\bm r}}\int {\rm d}^2{\bm q}{\rm e}^{{\rm i}{\bm q}\cdot{\bm r}}\Bigg[-\frac{M({\bm q})}{2{\rm i}}\big({\rm e}^{-{\rm i}{\bm k}\cdot{\bm r}}-{\rm e}^{{\rm i}{\bm k}\cdot{\bm r}}\big)+\frac{S({\bm q})}{2}\big({\rm e}^{-{\rm i}{\bm k}\cdot{\bm r}}+{\rm e}^{{\rm i}{\bm k}\cdot{\bm r}}\big) \\
   -\frac{a}{\ell}\frac{K_1({\bm q})}{{\rm i}}\big({\rm e}^{-2{\rm i}{\bm k}\cdot{\bm r}}-{\rm e}^{2{\rm i}{\bm k}\cdot{\bm r}}\big)+\frac{a}{\ell}J_1({\bm q})\big({\rm e}^{-2{\rm i}{\bm k}\cdot{\bm r}}+{\rm e}^{2{\rm i}{\bm k}\cdot{\bm r}}\big) \Bigg]. \tag{\theequation}
\end{align*}
Using the identity
\begin{equation}
   \sum_{{\bm r}}{\rm e}^{{\rm i}{\bm q}'\cdot{\bm r}}=\sum_{\bm G}\frac{4\pi}{\ell^2}\delta^2({\bm q}'-{\bm G})
\end{equation}
\stepcounter{equation}
we can write this as a sum over reciprocal lattice vectors, ${\bm G}$. Performing the ${\bm q}$ integral using the Dirac delta functions gives
\begin{align*}
   \tilde{\gamma}_{\bm k}({\bm q}')=\frac{\Omega_0 b \pi}{\ell^2}\sum_{{\bm G}}\Bigg[-\frac{M(-{\bm q}'+{\bm k}+{\bm G})-M(-{\bm q}'-{\bm k}+{\bm G})}{2{\rm i}}+\frac{S(-{\bm q}'+{\bm k}+{\bm G})+S(-{\bm q}'-{\bm k}+{\bm G})}{2}\\
   -\frac{a}{\ell}\frac{K_1(-{\bm q}'+2{\bm k}+{\bm G})-K(-{\bm q}'-2{\bm k}+{\bm G})}{{\rm i}}+\frac{a}{\ell}\big(J_1(-{\bm q}'+2{\bm k}+{\bm G})+J_1(-{\bm q}'-2{\bm k}+{\bm G})\big) \Bigg]. \tag{\theequation}
\end{align*}

\section{Exploring parameters: more dispersion and stability plots}
\begin{figure} 
   \centering
   \begin{tabular}{lll}
   (A) & (B) & (C) \\
   \includegraphics[width=0.3\textwidth]{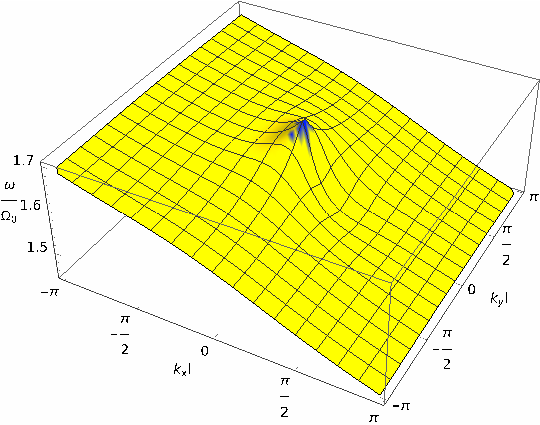} 
   &
   \includegraphics[width=0.3\textwidth]{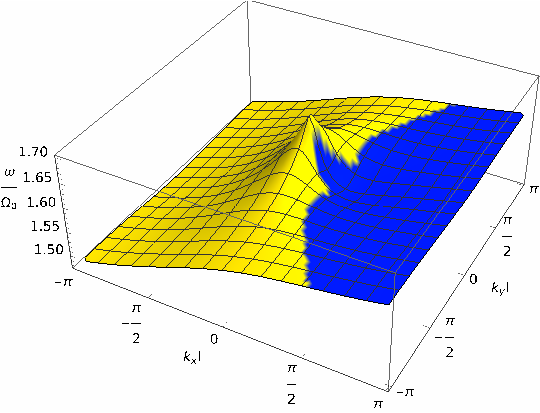} 
   &
   \includegraphics[width=0.3\textwidth]{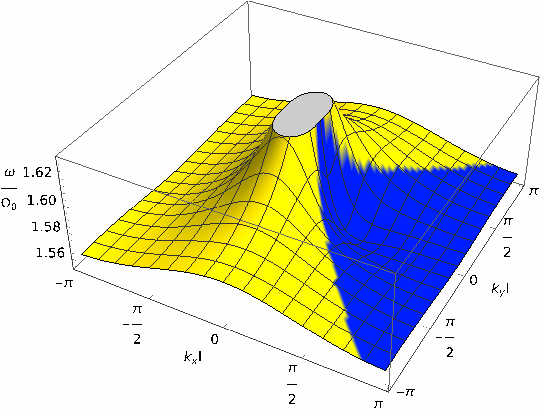} 
   \\
   (D) & (E) & (F) \\
   \includegraphics[width=0.3\textwidth]{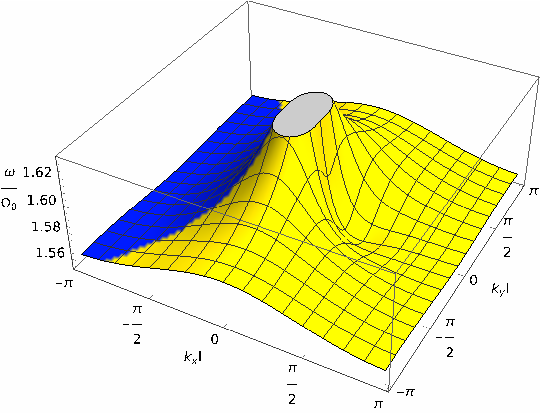} 
   &
   \includegraphics[width=0.3\textwidth]{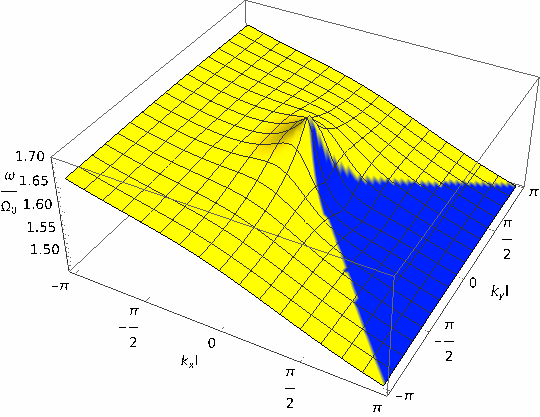} 
   &
   \includegraphics[width=0.3\textwidth]{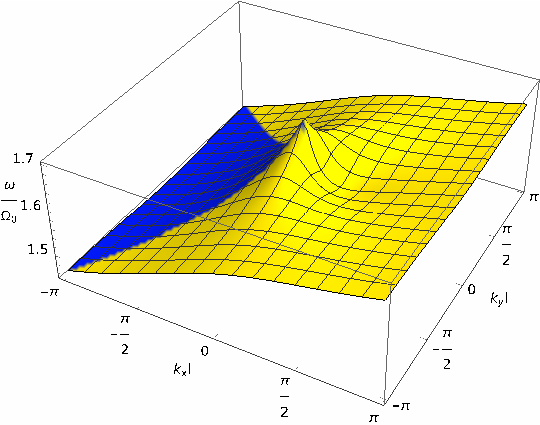} \\ 
   \end{tabular}
   \caption{Additional plots of dispersion relation and linear stability of metachronal waves using a square lattice. Linear stability analysis is represented by color: yellow regions are unstable and blue regions are stable. Parameters used for all plots are $a/\ell=0.3$, $b/\ell=0.03$, $h=\ell$, $\theta=0, \chi=0$. Parameters for individuals plots are: (A) $A_1=-0.3, B_1=0.4, A_2=0, B_2=0, D_1=-0.1, C_1=C_2=D_2=0$; (B) $A_1=0.3, B_1=-0.4, A_2=0, B_2=0, D_1=-0.1, C_1=C_2=D_2=0$; (C) $A_1=0.3, B_1=0, A_2=0, B_2=0, C_1=D_1=C_2=D_2=0$; (D) $A_1=-0.3, B_1=0, A_2=0, B_2=0, C_1=D_1=C_2=D_2=0$; (E) $A_1=0, B_1=0.4, A_2=0, B_2=0, C_1=D_1=C_2=D_2=0$; (F) $A_1=0.3, B_1=-0.4, A_2=0, B_2=0, C_1=D_1=C_2=D_2=0$.}
   \label{fig:extraplots}
\end{figure}

In Fig. \ref{fig:extraplots}, the dispersion and stability relation is shown for other parameter combinations that are not shown in the main text.
This demonstrates further how the first harmonic coefficients have a significant effect on stability. In Fig. \ref{fig:extraplots}(A), we do not see the usual stability region for a forwards or backwards wave. In Fig. \ref{fig:extraplots}(B), we see a stable forwards wave, but the shape of the stable region is different compared with those in the main text where $A_1$ and $B_1$ have the same sign. Another interesting feature of this plot is that the asymmetry of the dispersion relation shows a higher frequency in the stable wave direction than in the unstable direction, whereas for other plots shown when $A_1$ and $B_1$ have the same sign, we see that the asymmetry of the dispersion relation gives a lower frequency in the stable direction. When $B_1$ is zero, we see that the sign of $A_1$ gives stability regions in the expected direction, but the dispersion relation does not show a strong change in magnitude between the forwards and backwards direction (although there is a noticeable dip in the forwards direction just above the peak at small ${\bm k}$ which is also present in main text Fig. 2C when $A_1=B_1=0$). When $A_1=0$, the sign of $B_1$ gives the expected direction for the stability region and the expected asymmetry in the dispersion relation between forwards and backwards waves. This indicates that both $A_1$ and $B_1$ have an important effect on the direction of stability, but that the asymmetry in the dispersion relation, where one direction has a wave frequency than the other direction, is affected most significantly by $B_1$.


\begin{thebibliography}{59}%
\makeatletter
\providecommand \@ifxundefined [1]{%
 \@ifx{#1\undefined}
}%
\providecommand \@ifnum [1]{%
 \ifnum #1\expandafter \@firstoftwo
 \else \expandafter \@secondoftwo
 \fi
}%
\providecommand \@ifx [1]{%
 \ifx #1\expandafter \@firstoftwo
 \else \expandafter \@secondoftwo
 \fi
}%
\providecommand \natexlab [1]{#1}%
\providecommand \enquote  [1]{``#1''}%
\providecommand \bibnamefont  [1]{#1}%
\providecommand \bibfnamefont [1]{#1}%
\providecommand \citenamefont [1]{#1}%
\providecommand \href@noop [0]{\@secondoftwo}%
\providecommand \href [0]{\begingroup \@sanitize@url \@href}%
\providecommand \@href[1]{\@@startlink{#1}\@@href}%
\providecommand \@@href[1]{\endgroup#1\@@endlink}%
\providecommand \@sanitize@url [0]{\catcode `\\12\catcode `\$12\catcode
  `\&12\catcode `\#12\catcode `\^12\catcode `\_12\catcode `\%12\relax}%
\providecommand \@@startlink[1]{}%
\providecommand \@@endlink[0]{}%
\providecommand \url  [0]{\begingroup\@sanitize@url \@url }%
\providecommand \@url [1]{\endgroup\@href {#1}{\urlprefix }}%
\providecommand \urlprefix  [0]{URL }%
\providecommand \Eprint [0]{\href }%
\providecommand \doibase [0]{https://doi.org/}%
\providecommand \selectlanguage [0]{\@gobble}%
\providecommand \bibinfo  [0]{\@secondoftwo}%
\providecommand \bibfield  [0]{\@secondoftwo}%
\providecommand \translation [1]{[#1]}%
\providecommand \BibitemOpen [0]{}%
\providecommand \bibitemStop [0]{}%
\providecommand \bibitemNoStop [0]{.\EOS\space}%
\providecommand \EOS [0]{\spacefactor3000\relax}%
\providecommand \BibitemShut  [1]{\csname bibitem#1\endcsname}%
\let\auto@bib@innerbib\@empty
\bibitem [{\citenamefont {Gilpin}\ \emph {et~al.}(2020)\citenamefont {Gilpin},
  \citenamefont {Storm~Bull},\ and\ \citenamefont {Prakash}}]{Gilpin:2020}%
  \BibitemOpen
  \bibfield  {author} {\bibinfo {author} {\bibfnamefont {W.}~\bibnamefont
  {Gilpin}}, \bibinfo {author} {\bibfnamefont {M.}~\bibnamefont {Storm~Bull}},\
  and\ \bibinfo {author} {\bibfnamefont {M.}~\bibnamefont {Prakash}},\
  }\bibfield  {title} {\bibinfo {title} {The multiscale physics of cilia and
  flagella},\ }\href
  {https://doi.org/https://doi.org/10.1038/s42254-019-0129-0} {\bibfield
  {journal} {\bibinfo  {journal} {Nat. Rev. Phys.}\ }\textbf {\bibinfo {volume}
  {2}},\ \bibinfo {pages} {72} (\bibinfo {year} {2020})}\BibitemShut {NoStop}%
\bibitem [{\citenamefont {Gueron}\ and\ \citenamefont
  {Levit-Gurevich}(1999)}]{Gueron:1999}%
  \BibitemOpen
  \bibfield  {author} {\bibinfo {author} {\bibfnamefont {S.}~\bibnamefont
  {Gueron}}\ and\ \bibinfo {author} {\bibfnamefont {K.}~\bibnamefont
  {Levit-Gurevich}},\ }\bibfield  {title} {\bibinfo {title} {Energetic
  considerations of ciliary beating and the advantage of metachronal
  coordination},\ }\href {https://doi.org/10.1073/pnas.96.22.12240} {\bibfield
  {journal} {\bibinfo  {journal} {Proceedings of the National Academy of
  Sciences}\ }\textbf {\bibinfo {volume} {96}},\ \bibinfo {pages} {12240}
  (\bibinfo {year} {1999})}\BibitemShut {NoStop}%
\bibitem [{\citenamefont {Kim}\ and\ \citenamefont {Netz}(2006)}]{Kim2006}%
  \BibitemOpen
  \bibfield  {author} {\bibinfo {author} {\bibfnamefont {Y.~W.}\ \bibnamefont
  {Kim}}\ and\ \bibinfo {author} {\bibfnamefont {R.~R.}\ \bibnamefont {Netz}},\
  }\bibfield  {title} {\bibinfo {title} {{Pumping Fluids with Periodically
  Beating Grafted Elastic Filaments}},\ }\href@noop {} {\bibfield  {journal}
  {\bibinfo  {journal} {Physical Review Letters}\ }\textbf {\bibinfo {volume}
  {96}},\ \bibinfo {pages} {158101} (\bibinfo {year} {2006})}\BibitemShut
  {NoStop}%
\bibitem [{\citenamefont {Osterman}\ and\ \citenamefont
  {Vilfan}(2011)}]{Osterman2011}%
  \BibitemOpen
  \bibfield  {author} {\bibinfo {author} {\bibfnamefont {N.}~\bibnamefont
  {Osterman}}\ and\ \bibinfo {author} {\bibfnamefont {A.}~\bibnamefont
  {Vilfan}},\ }\bibfield  {title} {\bibinfo {title} {{Finding the ciliary
  beating pattern with optimal efficiency}},\ }\href@noop {} {\bibfield
  {journal} {\bibinfo  {journal} {Proceedings of the National Academy of
  Sciences}\ }\textbf {\bibinfo {volume} {108}},\ \bibinfo {pages} {15727}
  (\bibinfo {year} {2011})}\BibitemShut {NoStop}%
\bibitem [{\citenamefont {Elgeti}\ and\ \citenamefont
  {Gompper}(2013)}]{Elgeti2013}%
  \BibitemOpen
  \bibfield  {author} {\bibinfo {author} {\bibfnamefont {J.}~\bibnamefont
  {Elgeti}}\ and\ \bibinfo {author} {\bibfnamefont {G.}~\bibnamefont
  {Gompper}},\ }\bibfield  {title} {\bibinfo {title} {{Emergence of metachronal
  waves in cilia arrays}},\ }\href@noop {} {\bibfield  {journal} {\bibinfo
  {journal} {Proceedings of the National Academy of Sciences}\ }\textbf
  {\bibinfo {volume} {110}},\ \bibinfo {pages} {4470} (\bibinfo {year}
  {2013})}\BibitemShut {NoStop}%
\bibitem [{\citenamefont {Chateau}\ \emph {et~al.}(2017)\citenamefont
  {Chateau}, \citenamefont {Favier}, \citenamefont {D'Ortona},\ and\
  \citenamefont {Poncet}}]{Chateau2017}%
  \BibitemOpen
  \bibfield  {author} {\bibinfo {author} {\bibfnamefont {S.}~\bibnamefont
  {Chateau}}, \bibinfo {author} {\bibfnamefont {J.}~\bibnamefont {Favier}},
  \bibinfo {author} {\bibfnamefont {U.}~\bibnamefont {D'Ortona}},\ and\
  \bibinfo {author} {\bibfnamefont {S.}~\bibnamefont {Poncet}},\ }\bibfield
  {title} {\bibinfo {title} {{Transport efficiency of metachronal waves in 3D
  cilium arrays immersed in a two-phase flow}},\ }\href@noop {} {\bibfield
  {journal} {\bibinfo  {journal} {Journal of Fluid Mechanics}\ }\textbf
  {\bibinfo {volume} {824}},\ \bibinfo {pages} {931} (\bibinfo {year}
  {2017})}\BibitemShut {NoStop}%
\bibitem [{\citenamefont {Fahy}\ and\ \citenamefont
  {Dickey}(2010)}]{Fahy:2010}%
  \BibitemOpen
  \bibfield  {author} {\bibinfo {author} {\bibfnamefont {J.~V.}\ \bibnamefont
  {Fahy}}\ and\ \bibinfo {author} {\bibfnamefont {B.~F.}\ \bibnamefont
  {Dickey}},\ }\bibfield  {title} {\bibinfo {title} {Airway mucus function and
  dysfunction},\ }\href {https://doi.org/10.1056/NEJMra0910061} {\bibfield
  {journal} {\bibinfo  {journal} {New England Journal of Medicine}\ }\textbf
  {\bibinfo {volume} {363}},\ \bibinfo {pages} {2233} (\bibinfo {year}
  {2010})},\ \bibinfo {note} {pMID: 21121836}\BibitemShut {NoStop}%
\bibitem [{\citenamefont {Yaghi}\ and\ \citenamefont
  {Dolovich}(2016)}]{Yaghi:2016}%
  \BibitemOpen
  \bibfield  {author} {\bibinfo {author} {\bibfnamefont {A.}~\bibnamefont
  {Yaghi}}\ and\ \bibinfo {author} {\bibfnamefont {M.~B.}\ \bibnamefont
  {Dolovich}},\ }\bibfield  {title} {\bibinfo {title} {Airway epithelial cell
  cilia and obstructive lung disease},\ }\bibfield  {journal} {\bibinfo
  {journal} {Cells}\ }\textbf {\bibinfo {volume} {5}},\ \href
  {https://doi.org/10.3390/cells5040040} {10.3390/cells5040040} (\bibinfo
  {year} {2016})\BibitemShut {NoStop}%
\bibitem [{\citenamefont {Yuan}\ \emph {et~al.}(2021)\citenamefont {Yuan},
  \citenamefont {Wang}, \citenamefont {Peng}, \citenamefont {Ward},
  \citenamefont {Hennig}, \citenamefont {Zheng},\ and\ \citenamefont
  {Yan}}]{Yuan:2021}%
  \BibitemOpen
  \bibfield  {author} {\bibinfo {author} {\bibfnamefont {S.}~\bibnamefont
  {Yuan}}, \bibinfo {author} {\bibfnamefont {Z.}~\bibnamefont {Wang}}, \bibinfo
  {author} {\bibfnamefont {H.}~\bibnamefont {Peng}}, \bibinfo {author}
  {\bibfnamefont {S.~M.}\ \bibnamefont {Ward}}, \bibinfo {author}
  {\bibfnamefont {G.~W.}\ \bibnamefont {Hennig}}, \bibinfo {author}
  {\bibfnamefont {H.}~\bibnamefont {Zheng}},\ and\ \bibinfo {author}
  {\bibfnamefont {W.}~\bibnamefont {Yan}},\ }\bibfield  {title} {\bibinfo
  {title} {Oviductal motile cilia are essential for oocyte pickup but
  dispensable for sperm and embryo transport},\ }\href
  {https://doi.org/10.1073/pnas.2102940118} {\bibfield  {journal} {\bibinfo
  {journal} {Proceedings of the National Academy of Sciences}\ }\textbf
  {\bibinfo {volume} {118}},\ \bibinfo {pages} {e2102940118} (\bibinfo {year}
  {2021})}\BibitemShut {NoStop}%
\bibitem [{\citenamefont {Hirokawa}\ \emph {et~al.}(2009)\citenamefont
  {Hirokawa}, \citenamefont {Okada},\ and\ \citenamefont
  {Tanaka}}]{Hirokawa:2009}%
  \BibitemOpen
  \bibfield  {author} {\bibinfo {author} {\bibfnamefont {N.}~\bibnamefont
  {Hirokawa}}, \bibinfo {author} {\bibfnamefont {Y.}~\bibnamefont {Okada}},\
  and\ \bibinfo {author} {\bibfnamefont {Y.}~\bibnamefont {Tanaka}},\
  }\bibfield  {title} {\bibinfo {title} {Fluid dynamic mechanism responsible
  for breaking the left-right symmetry of the human body: The nodal flow},\
  }\href {https://doi.org/10.1146/annurev.fluid.010908.165141} {\bibfield
  {journal} {\bibinfo  {journal} {Annual Review of Fluid Mechanics}\ }\textbf
  {\bibinfo {volume} {41}},\ \bibinfo {pages} {53} (\bibinfo {year}
  {2009})}\BibitemShut {NoStop}%
\bibitem [{\citenamefont {Smith}\ \emph {et~al.}(2019)\citenamefont {Smith},
  \citenamefont {Montenegro-Johnson},\ and\ \citenamefont
  {Lopes}}]{Smith:2019}%
  \BibitemOpen
  \bibfield  {author} {\bibinfo {author} {\bibfnamefont {D.~J.}\ \bibnamefont
  {Smith}}, \bibinfo {author} {\bibfnamefont {T.~D.}\ \bibnamefont
  {Montenegro-Johnson}},\ and\ \bibinfo {author} {\bibfnamefont {S.~S.}\
  \bibnamefont {Lopes}},\ }\bibfield  {title} {\bibinfo {title}
  {Symmetry-breaking cilia-driven flow in embryogenesis},\ }\href
  {https://doi.org/10.1146/annurev-fluid-010518-040231} {\bibfield  {journal}
  {\bibinfo  {journal} {Annual Review of Fluid Mechanics}\ }\textbf {\bibinfo
  {volume} {51}},\ \bibinfo {pages} {105} (\bibinfo {year} {2019})}\BibitemShut
  {NoStop}%
\bibitem [{\citenamefont {Faubel}\ \emph {et~al.}(2016)\citenamefont {Faubel},
  \citenamefont {Westendorf}, \citenamefont {Bodenschatz},\ and\ \citenamefont
  {Eichele}}]{Faubel:2016}%
  \BibitemOpen
  \bibfield  {author} {\bibinfo {author} {\bibfnamefont {R.}~\bibnamefont
  {Faubel}}, \bibinfo {author} {\bibfnamefont {C.}~\bibnamefont {Westendorf}},
  \bibinfo {author} {\bibfnamefont {E.}~\bibnamefont {Bodenschatz}},\ and\
  \bibinfo {author} {\bibfnamefont {G.}~\bibnamefont {Eichele}},\ }\bibfield
  {title} {\bibinfo {title} {Cilia-based flow network in the brain
  ventricles},\ }\href {https://doi.org/10.1126/science.aae0450} {\bibfield
  {journal} {\bibinfo  {journal} {Science}\ }\textbf {\bibinfo {volume}
  {353}},\ \bibinfo {pages} {176} (\bibinfo {year} {2016})}\BibitemShut
  {NoStop}%
\bibitem [{\citenamefont {Pellicciotta}\ \emph {et~al.}(2020)\citenamefont
  {Pellicciotta}, \citenamefont {Hamilton}, \citenamefont {Kotar},
  \citenamefont {Faucourt}, \citenamefont {Delgehyr}, \citenamefont {Spassky},\
  and\ \citenamefont {Cicuta}}]{Pellicciotta:2020}%
  \BibitemOpen
  \bibfield  {author} {\bibinfo {author} {\bibfnamefont {N.}~\bibnamefont
  {Pellicciotta}}, \bibinfo {author} {\bibfnamefont {E.}~\bibnamefont
  {Hamilton}}, \bibinfo {author} {\bibfnamefont {J.}~\bibnamefont {Kotar}},
  \bibinfo {author} {\bibfnamefont {M.}~\bibnamefont {Faucourt}}, \bibinfo
  {author} {\bibfnamefont {N.}~\bibnamefont {Delgehyr}}, \bibinfo {author}
  {\bibfnamefont {N.}~\bibnamefont {Spassky}},\ and\ \bibinfo {author}
  {\bibfnamefont {P.}~\bibnamefont {Cicuta}},\ }\bibfield  {title} {\bibinfo
  {title} {Entrainment of mammalian motile cilia in the brain with hydrodynamic
  forces},\ }\href@noop {} {\bibfield  {journal} {\bibinfo  {journal} {Proc.
  Natl. Acad. Sci. U.S.A.}\ }\textbf {\bibinfo {volume} {117}},\ \bibinfo
  {pages} {8315} (\bibinfo {year} {2020})}\BibitemShut {NoStop}%
\bibitem [{\citenamefont {Brumley}\ \emph {et~al.}(2012)\citenamefont
  {Brumley}, \citenamefont {Polin}, \citenamefont {Pedley},\ and\ \citenamefont
  {Goldstein}}]{Brumley2012}%
  \BibitemOpen
  \bibfield  {author} {\bibinfo {author} {\bibfnamefont {D.~R.}\ \bibnamefont
  {Brumley}}, \bibinfo {author} {\bibfnamefont {M.}~\bibnamefont {Polin}},
  \bibinfo {author} {\bibfnamefont {T.~J.}\ \bibnamefont {Pedley}},\ and\
  \bibinfo {author} {\bibfnamefont {R.~E.}\ \bibnamefont {Goldstein}},\
  }\bibfield  {title} {\bibinfo {title} {{Hydrodynamic Synchronization and
  Metachronal Waves on the Surface of the Colonial Alga Volvox carteri}},\
  }\href@noop {} {\bibfield  {journal} {\bibinfo  {journal} {Physical Review
  Letters}\ }\textbf {\bibinfo {volume} {109}},\ \bibinfo {pages} {268102}
  (\bibinfo {year} {2012})}\BibitemShut {NoStop}%
\bibitem [{\citenamefont {Feriani}\ \emph {et~al.}(2017)\citenamefont
  {Feriani}, \citenamefont {Juenet}, \citenamefont {Fowler}, \citenamefont
  {Bruot}, \citenamefont {Chiocciolo}, \citenamefont {Holland}, \citenamefont
  {Bryant},\ and\ \citenamefont {Cicuta}}]{Feriani:2017}%
  \BibitemOpen
  \bibfield  {author} {\bibinfo {author} {\bibfnamefont {L.}~\bibnamefont
  {Feriani}}, \bibinfo {author} {\bibfnamefont {M.}~\bibnamefont {Juenet}},
  \bibinfo {author} {\bibfnamefont {C.~J.}\ \bibnamefont {Fowler}}, \bibinfo
  {author} {\bibfnamefont {N.}~\bibnamefont {Bruot}}, \bibinfo {author}
  {\bibfnamefont {M.}~\bibnamefont {Chiocciolo}}, \bibinfo {author}
  {\bibfnamefont {S.~M.}\ \bibnamefont {Holland}}, \bibinfo {author}
  {\bibfnamefont {C.~E.}\ \bibnamefont {Bryant}},\ and\ \bibinfo {author}
  {\bibfnamefont {P.}~\bibnamefont {Cicuta}},\ }\bibfield  {title} {\bibinfo
  {title} {Assessing the collective dynamics of motile cilia in cultures of
  human airway cells by multiscale {DDM}},\ }\href@noop {} {\bibfield
  {journal} {\bibinfo  {journal} {Biophysical Journal}\ }\textbf {\bibinfo
  {volume} {113}},\ \bibinfo {pages} {109} (\bibinfo {year}
  {2017})}\BibitemShut {NoStop}%
\bibitem [{\citenamefont {Chioccioli}\ \emph {et~al.}(2019)\citenamefont
  {Chioccioli}, \citenamefont {Feriani}, \citenamefont {Kotar}, \citenamefont
  {Bratcher},\ and\ \citenamefont {Cicuta}}]{Chiocciolo:2019}%
  \BibitemOpen
  \bibfield  {author} {\bibinfo {author} {\bibfnamefont {M.}~\bibnamefont
  {Chioccioli}}, \bibinfo {author} {\bibfnamefont {L.}~\bibnamefont {Feriani}},
  \bibinfo {author} {\bibfnamefont {J.}~\bibnamefont {Kotar}}, \bibinfo
  {author} {\bibfnamefont {P.}~\bibnamefont {Bratcher}},\ and\ \bibinfo
  {author} {\bibfnamefont {P.}~\bibnamefont {Cicuta}},\ }\bibfield  {title}
  {\bibinfo {title} {Phenotyping ciliary dynamics and coordination in response
  to {CFTR-}modulators in cystic fibrosis respiratory epithelial cells},\
  }\href {https://doi.org/10.1038/s41467-019-09798-3} {\bibfield  {journal}
  {\bibinfo  {journal} {Nat. Commun.}\ }\textbf {\bibinfo {volume} {10}},\
  \bibinfo {pages} {1763} (\bibinfo {year} {2019})}\BibitemShut {NoStop}%
\bibitem [{\citenamefont {Burn}\ \emph {et~al.}(2022)\citenamefont {Burn},
  \citenamefont {Schneiter}, \citenamefont {Ryser}, \citenamefont {Gehr},
  \citenamefont {Ri\u{c}ka},\ and\ \citenamefont {Frenz}}]{Burn:2022}%
  \BibitemOpen
  \bibfield  {author} {\bibinfo {author} {\bibfnamefont {A.}~\bibnamefont
  {Burn}}, \bibinfo {author} {\bibfnamefont {M.}~\bibnamefont {Schneiter}},
  \bibinfo {author} {\bibfnamefont {M.}~\bibnamefont {Ryser}}, \bibinfo
  {author} {\bibfnamefont {P.}~\bibnamefont {Gehr}}, \bibinfo {author}
  {\bibfnamefont {J.}~\bibnamefont {Ri\u{c}ka}},\ and\ \bibinfo {author}
  {\bibfnamefont {M.}~\bibnamefont {Frenz}},\ }\bibfield  {title} {\bibinfo
  {title} {A quantitative interspecies comparison of the respiratory
  mucociliary clearance mechanism},\ }\href
  {https://doi.org/10.1007/s00249-021-01584-8} {\bibfield  {journal} {\bibinfo
  {journal} {Eur Biophys J}\ }\textbf {\bibinfo {volume} {51}},\ \bibinfo
  {pages} {51} (\bibinfo {year} {2022})}\BibitemShut {NoStop}%
\bibitem [{\citenamefont {Poon}\ \emph {et~al.}(2022)\citenamefont {Poon},
  \citenamefont {Westwood}, \citenamefont {Laeverenz-Schlogelhofer},
  \citenamefont {Brodrick}, \citenamefont {Craggs}, \citenamefont {Keaveny},
  \citenamefont {J{\'e}kely},\ and\ \citenamefont {Wan}}]{Poon:2022}%
  \BibitemOpen
  \bibfield  {author} {\bibinfo {author} {\bibfnamefont {R.~N.}\ \bibnamefont
  {Poon}}, \bibinfo {author} {\bibfnamefont {T.~A.}\ \bibnamefont {Westwood}},
  \bibinfo {author} {\bibfnamefont {H.}~\bibnamefont
  {Laeverenz-Schlogelhofer}}, \bibinfo {author} {\bibfnamefont
  {E.}~\bibnamefont {Brodrick}}, \bibinfo {author} {\bibfnamefont
  {J.}~\bibnamefont {Craggs}}, \bibinfo {author} {\bibfnamefont {E.~E.}\
  \bibnamefont {Keaveny}}, \bibinfo {author} {\bibfnamefont {G.}~\bibnamefont
  {J{\'e}kely}},\ and\ \bibinfo {author} {\bibfnamefont {K.~Y.}\ \bibnamefont
  {Wan}},\ }\bibfield  {title} {\bibinfo {title} {Ciliary propulsion and
  metachronal coordination in reef coral larvae},\ }\bibfield  {journal}
  {\bibinfo  {journal} {bioRxiv}\ }\href
  {https://doi.org/10.1101/2022.09.19.508546} {10.1101/2022.09.19.508546}
  (\bibinfo {year} {2022})\BibitemShut {NoStop}%
\bibitem [{\citenamefont {Ringers}\ \emph {et~al.}(2023)\citenamefont
  {Ringers}, \citenamefont {Bialonski}, \citenamefont {Ege}, \citenamefont
  {Solovev}, \citenamefont {Hansen}, \citenamefont {Jeong}, \citenamefont
  {Friedrich},\ and\ \citenamefont {Jurisch-Yaksi}}]{Ringers:2023}%
  \BibitemOpen
  \bibfield  {author} {\bibinfo {author} {\bibfnamefont {C.}~\bibnamefont
  {Ringers}}, \bibinfo {author} {\bibfnamefont {S.}~\bibnamefont {Bialonski}},
  \bibinfo {author} {\bibfnamefont {M.}~\bibnamefont {Ege}}, \bibinfo {author}
  {\bibfnamefont {A.}~\bibnamefont {Solovev}}, \bibinfo {author} {\bibfnamefont
  {J.~N.}\ \bibnamefont {Hansen}}, \bibinfo {author} {\bibfnamefont
  {I.}~\bibnamefont {Jeong}}, \bibinfo {author} {\bibfnamefont {B.~M.}\
  \bibnamefont {Friedrich}},\ and\ \bibinfo {author} {\bibfnamefont
  {N.}~\bibnamefont {Jurisch-Yaksi}},\ }\bibfield  {title} {\bibinfo {title}
  {Novel analytical tools reveal that local synchronization of cilia coincides
  with tissue-scale metachronal waves in zebrafish multiciliated epithelia},\
  }\href {https://doi.org/10.7554/eLife.77701} {\bibfield  {journal} {\bibinfo
  {journal} {eLife}\ }\textbf {\bibinfo {volume} {12}},\ \bibinfo {pages}
  {e77701} (\bibinfo {year} {2023})}\BibitemShut {NoStop}%
\bibitem [{\citenamefont {Machemer}(1972)}]{Machemer:1972}%
  \BibitemOpen
  \bibfield  {author} {\bibinfo {author} {\bibfnamefont {H.}~\bibnamefont
  {Machemer}},\ }\bibfield  {title} {\bibinfo {title} {Ciliary activity and the
  origin of metachrony in {\it paramecium}: Effects of increased viscosity},\
  }\href@noop {} {\bibfield  {journal} {\bibinfo  {journal} {J Exp Biol}\
  }\textbf {\bibinfo {volume} {57}},\ \bibinfo {pages} {239} (\bibinfo {year}
  {1972})}\BibitemShut {NoStop}%
\bibitem [{\citenamefont {Cicuta}(2020)}]{Cicuta:2020}%
  \BibitemOpen
  \bibfield  {author} {\bibinfo {author} {\bibfnamefont {P.}~\bibnamefont
  {Cicuta}},\ }\bibfield  {title} {\bibinfo {title} {{The use of biophysical
  approaches to understand ciliary beating}},\ }\href
  {https://doi.org/10.1042/BST20190571} {\bibfield  {journal} {\bibinfo
  {journal} {Biochemical Society Transactions}\ }\textbf {\bibinfo {volume}
  {48}},\ \bibinfo {pages} {221} (\bibinfo {year} {2020})}\BibitemShut
  {NoStop}%
\bibitem [{\citenamefont {Guirao}\ and\ \citenamefont
  {Joanny}(2007)}]{Guirao2007}%
  \BibitemOpen
  \bibfield  {author} {\bibinfo {author} {\bibfnamefont {B.}~\bibnamefont
  {Guirao}}\ and\ \bibinfo {author} {\bibfnamefont {J.-F.}\ \bibnamefont
  {Joanny}},\ }\bibfield  {title} {\bibinfo {title} {{Spontaneous Creation of
  Macroscopic Flow and Metachronal Waves in an Array of Cilia}},\ }\href@noop
  {} {\bibfield  {journal} {\bibinfo  {journal} {Biophysical Journal}\ }\textbf
  {\bibinfo {volume} {92}},\ \bibinfo {pages} {1900} (\bibinfo {year}
  {2007})}\BibitemShut {NoStop}%
\bibitem [{\citenamefont {Wollin}\ and\ \citenamefont
  {Stark}(2011)}]{Wollin2011}%
  \BibitemOpen
  \bibfield  {author} {\bibinfo {author} {\bibfnamefont {C.}~\bibnamefont
  {Wollin}}\ and\ \bibinfo {author} {\bibfnamefont {H.}~\bibnamefont {Stark}},\
  }\bibfield  {title} {\bibinfo {title} {{Metachronal waves in a chain of
  rowers with hydrodynamic interactions}},\ }\href@noop {} {\bibfield
  {journal} {\bibinfo  {journal} {The European Physical Journal E}\ }\textbf
  {\bibinfo {volume} {34}},\ \bibinfo {pages} {42} (\bibinfo {year}
  {2011})}\BibitemShut {NoStop}%
\bibitem [{\citenamefont {Brumley}\ \emph {et~al.}(2015)\citenamefont
  {Brumley}, \citenamefont {Polin}, \citenamefont {Pedley},\ and\ \citenamefont
  {Goldstein}}]{Brumley2015}%
  \BibitemOpen
  \bibfield  {author} {\bibinfo {author} {\bibfnamefont {D.~R.}\ \bibnamefont
  {Brumley}}, \bibinfo {author} {\bibfnamefont {M.}~\bibnamefont {Polin}},
  \bibinfo {author} {\bibfnamefont {T.~J.}\ \bibnamefont {Pedley}},\ and\
  \bibinfo {author} {\bibfnamefont {R.~E.}\ \bibnamefont {Goldstein}},\
  }\bibfield  {title} {\bibinfo {title} {{Metachronal waves in the flagellar
  beating of Volvox and their hydrodynamic origin}},\ }\href@noop {} {\bibfield
   {journal} {\bibinfo  {journal} {J. R. Soc. Interface}\ }\textbf {\bibinfo
  {volume} {12}},\ \bibinfo {pages} {20141358} (\bibinfo {year}
  {2015})}\BibitemShut {NoStop}%
\bibitem [{\citenamefont {Ghorbani}\ and\ \citenamefont
  {Najafi}(2017)}]{Ghorbani2017}%
  \BibitemOpen
  \bibfield  {author} {\bibinfo {author} {\bibfnamefont {A.}~\bibnamefont
  {Ghorbani}}\ and\ \bibinfo {author} {\bibfnamefont {A.}~\bibnamefont
  {Najafi}},\ }\bibfield  {title} {\bibinfo {title} {{Symplectic and
  antiplectic waves in an array of beating cilia attached to a closed body}},\
  }\href@noop {} {\bibfield  {journal} {\bibinfo  {journal} {Physical Review
  E}\ }\textbf {\bibinfo {volume} {95}},\ \bibinfo {pages} {052412} (\bibinfo
  {year} {2017})}\BibitemShut {NoStop}%
\bibitem [{\citenamefont {Han}\ and\ \citenamefont {Peskin}(2018)}]{Han2018}%
  \BibitemOpen
  \bibfield  {author} {\bibinfo {author} {\bibfnamefont {J.}~\bibnamefont
  {Han}}\ and\ \bibinfo {author} {\bibfnamefont {C.~S.}\ \bibnamefont
  {Peskin}},\ }\bibfield  {title} {\bibinfo {title} {{Spontaneous oscillation
  and fluid-structure interaction of cilia}},\ }\href@noop {} {\bibfield
  {journal} {\bibinfo  {journal} {Proceedings of the National Academy of
  Sciences}\ }\textbf {\bibinfo {volume} {115}},\ \bibinfo {pages} {4417}
  (\bibinfo {year} {2018})}\BibitemShut {NoStop}%
\bibitem [{\citenamefont {Meng}\ \emph {et~al.}(2021)\citenamefont {Meng},
  \citenamefont {Bennett}, \citenamefont {Uchida},\ and\ \citenamefont
  {Golestanian}}]{Meng:2021}%
  \BibitemOpen
  \bibfield  {author} {\bibinfo {author} {\bibfnamefont {F.}~\bibnamefont
  {Meng}}, \bibinfo {author} {\bibfnamefont {R.~R.}\ \bibnamefont {Bennett}},
  \bibinfo {author} {\bibfnamefont {N.}~\bibnamefont {Uchida}},\ and\ \bibinfo
  {author} {\bibfnamefont {R.}~\bibnamefont {Golestanian}},\ }\bibfield
  {title} {\bibinfo {title} {Conditions for metachronal coordination in arrays
  of model cilia},\ }\href@noop {} {\bibfield  {journal} {\bibinfo  {journal}
  {Proc. Natl. Acad. Sci.}\ }\textbf {\bibinfo {volume} {118}},\ \bibinfo
  {pages} {e2102828118} (\bibinfo {year} {2021})}\BibitemShut {NoStop}%
\bibitem [{\citenamefont {Westwood}\ and\ \citenamefont
  {Keaveny}(2021)}]{Westwood:2021}%
  \BibitemOpen
  \bibfield  {author} {\bibinfo {author} {\bibfnamefont {T.~A.}\ \bibnamefont
  {Westwood}}\ and\ \bibinfo {author} {\bibfnamefont {E.~E.}\ \bibnamefont
  {Keaveny}},\ }\bibfield  {title} {\bibinfo {title} {Coordinated motion of
  active filaments on spherical surfaces},\ }\href
  {https://doi.org/10.1103/PhysRevFluids.6.L121101} {\bibfield  {journal}
  {\bibinfo  {journal} {Phys. Rev. Fluids}\ }\textbf {\bibinfo {volume} {6}},\
  \bibinfo {pages} {L121101} (\bibinfo {year} {2021})}\BibitemShut {NoStop}%
\bibitem [{\citenamefont {Solovev}\ and\ \citenamefont
  {Friedrich}(2022)}]{Solovev:2022}%
  \BibitemOpen
  \bibfield  {author} {\bibinfo {author} {\bibfnamefont {A.}~\bibnamefont
  {Solovev}}\ and\ \bibinfo {author} {\bibfnamefont {B.~M.}\ \bibnamefont
  {Friedrich}},\ }\bibfield  {title} {\bibinfo {title} {Synchronization in
  cilia carpets: multiple metachronal waves are stable, but one wave
  dominates},\ }\href {https://doi.org/10.1088/1367-2630/ac2ae4} {\bibfield
  {journal} {\bibinfo  {journal} {New Journal of Physics}\ }\textbf {\bibinfo
  {volume} {24}},\ \bibinfo {pages} {013015} (\bibinfo {year}
  {2022})}\BibitemShut {NoStop}%
\bibitem [{\citenamefont {Kanale}\ \emph {et~al.}(2022)\citenamefont {Kanale},
  \citenamefont {Ling}, \citenamefont {Guo}, \citenamefont {F\"{u}rthauer},\
  and\ \citenamefont {Kanso}}]{Kanale:2022}%
  \BibitemOpen
  \bibfield  {author} {\bibinfo {author} {\bibfnamefont {A.~V.}\ \bibnamefont
  {Kanale}}, \bibinfo {author} {\bibfnamefont {F.}~\bibnamefont {Ling}},
  \bibinfo {author} {\bibfnamefont {H.}~\bibnamefont {Guo}}, \bibinfo {author}
  {\bibfnamefont {S.}~\bibnamefont {F\"{u}rthauer}},\ and\ \bibinfo {author}
  {\bibfnamefont {E.}~\bibnamefont {Kanso}},\ }\bibfield  {title} {\bibinfo
  {title} {Spontaneous phase coordination and fluid pumping in model ciliary
  carpets},\ }\href {https://doi.org/10.1073/pnas.2214413119} {\bibfield
  {journal} {\bibinfo  {journal} {Proceedings of the National Academy of
  Sciences}\ }\textbf {\bibinfo {volume} {119}},\ \bibinfo {pages}
  {e2214413119} (\bibinfo {year} {2022})}\BibitemShut {NoStop}%
\bibitem [{\citenamefont {Kotar}\ \emph {et~al.}(2010)\citenamefont {Kotar},
  \citenamefont {Leoni}, \citenamefont {Bassetti}, \citenamefont
  {Lagomarsino},\ and\ \citenamefont {Cicuta}}]{Kotar:2010}%
  \BibitemOpen
  \bibfield  {author} {\bibinfo {author} {\bibfnamefont {J.}~\bibnamefont
  {Kotar}}, \bibinfo {author} {\bibfnamefont {M.}~\bibnamefont {Leoni}},
  \bibinfo {author} {\bibfnamefont {B.}~\bibnamefont {Bassetti}}, \bibinfo
  {author} {\bibfnamefont {M.~C.}\ \bibnamefont {Lagomarsino}},\ and\ \bibinfo
  {author} {\bibfnamefont {P.}~\bibnamefont {Cicuta}},\ }\bibfield  {title}
  {\bibinfo {title} {Hydrodynamic synchronization of colloidal oscillators},\
  }\href {https://doi.org/10.1073/pnas.0912455107} {\bibfield  {journal}
  {\bibinfo  {journal} {Proceedings of the National Academy of Sciences}\
  }\textbf {\bibinfo {volume} {107}},\ \bibinfo {pages} {7669} (\bibinfo {year}
  {2010})}\BibitemShut {NoStop}%
\bibitem [{\citenamefont {Kavre}\ \emph {et~al.}(2015)\citenamefont {Kavre},
  \citenamefont {Vilfan},\ and\ \citenamefont {Babi\ifmmode~\check{c}\else
  \v{c}\fi{}}}]{Kavre:2015}%
  \BibitemOpen
  \bibfield  {author} {\bibinfo {author} {\bibfnamefont {I.}~\bibnamefont
  {Kavre}}, \bibinfo {author} {\bibfnamefont {A.}~\bibnamefont {Vilfan}},\ and\
  \bibinfo {author} {\bibfnamefont {D.~c.~v.}\ \bibnamefont
  {Babi\ifmmode~\check{c}\else \v{c}\fi{}}},\ }\bibfield  {title} {\bibinfo
  {title} {Hydrodynamic synchronization of autonomously oscillating optically
  trapped particles},\ }\href {https://doi.org/10.1103/PhysRevE.91.031002}
  {\bibfield  {journal} {\bibinfo  {journal} {Phys. Rev. E}\ }\textbf {\bibinfo
  {volume} {91}},\ \bibinfo {pages} {031002(R)} (\bibinfo {year}
  {2015})}\BibitemShut {NoStop}%
\bibitem [{\citenamefont {Maestro}\ \emph {et~al.}(2018)\citenamefont
  {Maestro}, \citenamefont {Bruot}, \citenamefont {Kotar}, \citenamefont
  {Uchida}, \citenamefont {Golestanian},\ and\ \citenamefont
  {Cicuta}}]{Maestro2018}%
  \BibitemOpen
  \bibfield  {author} {\bibinfo {author} {\bibfnamefont {A.}~\bibnamefont
  {Maestro}}, \bibinfo {author} {\bibfnamefont {N.}~\bibnamefont {Bruot}},
  \bibinfo {author} {\bibfnamefont {J.}~\bibnamefont {Kotar}}, \bibinfo
  {author} {\bibfnamefont {N.}~\bibnamefont {Uchida}}, \bibinfo {author}
  {\bibfnamefont {R.}~\bibnamefont {Golestanian}},\ and\ \bibinfo {author}
  {\bibfnamefont {P.}~\bibnamefont {Cicuta}},\ }\bibfield  {title} {\bibinfo
  {title} {{Control of synchronization in models of hydrodynamically coupled
  motile cilia}},\ }\href@noop {} {\bibfield  {journal} {\bibinfo  {journal}
  {Commun. Phys.}\ }\textbf {\bibinfo {volume} {1}},\ \bibinfo {pages} {28}
  (\bibinfo {year} {2018})}\BibitemShut {NoStop}%
\bibitem [{\citenamefont {Brumley}\ \emph {et~al.}(2014)\citenamefont
  {Brumley}, \citenamefont {Wan}, \citenamefont {Polin},\ and\ \citenamefont
  {Goldstein}}]{Brumley2014}%
  \BibitemOpen
  \bibfield  {author} {\bibinfo {author} {\bibfnamefont {D.~R.}\ \bibnamefont
  {Brumley}}, \bibinfo {author} {\bibfnamefont {K.~Y.}\ \bibnamefont {Wan}},
  \bibinfo {author} {\bibfnamefont {M.}~\bibnamefont {Polin}},\ and\ \bibinfo
  {author} {\bibfnamefont {R.~E.}\ \bibnamefont {Goldstein}},\ }\bibfield
  {title} {\bibinfo {title} {Flagellar synchronization through direct
  hydrodynamic interactions},\ }\href@noop {} {\bibfield  {journal} {\bibinfo
  {journal} {Elife}\ }\textbf {\bibinfo {volume} {3}},\ \bibinfo {pages}
  {e02750} (\bibinfo {year} {2014})}\BibitemShut {NoStop}%
\bibitem [{\citenamefont {Narematsu}\ \emph {et~al.}(2015)\citenamefont
  {Narematsu}, \citenamefont {Quek}, \citenamefont {Chiam},\ and\ \citenamefont
  {Iwadate}}]{Narematsu:2015}%
  \BibitemOpen
  \bibfield  {author} {\bibinfo {author} {\bibfnamefont {N.}~\bibnamefont
  {Narematsu}}, \bibinfo {author} {\bibfnamefont {R.}~\bibnamefont {Quek}},
  \bibinfo {author} {\bibfnamefont {K.-H.}\ \bibnamefont {Chiam}},\ and\
  \bibinfo {author} {\bibfnamefont {Y.}~\bibnamefont {Iwadate}},\ }\bibfield
  {title} {\bibinfo {title} {Ciliary metachronal wave propagation on the
  compliant surface of paramecium cells},\ }\href
  {https://doi.org/10.1002/cm.21266} {\bibfield  {journal} {\bibinfo  {journal}
  {Cytoskeleton}\ }\textbf {\bibinfo {volume} {72}},\ \bibinfo {pages} {633}
  (\bibinfo {year} {2015})}\BibitemShut {NoStop}%
\bibitem [{\citenamefont {Wan}\ and\ \citenamefont
  {Goldstein}(2016)}]{Wan2016}%
  \BibitemOpen
  \bibfield  {author} {\bibinfo {author} {\bibfnamefont {K.~Y.}\ \bibnamefont
  {Wan}}\ and\ \bibinfo {author} {\bibfnamefont {R.~E.}\ \bibnamefont
  {Goldstein}},\ }\bibfield  {title} {\bibinfo {title} {{Coordinated beating of
  algal flagella is mediated by basal coupling}},\ }\href@noop {} {\bibfield
  {journal} {\bibinfo  {journal} {Proc. Natl. Acad. Sci.}\ }\textbf {\bibinfo
  {volume} {113}},\ \bibinfo {pages} {E2784} (\bibinfo {year}
  {2016})}\BibitemShut {NoStop}%
\bibitem [{\citenamefont {Klindt}\ \emph {et~al.}(2017)\citenamefont {Klindt},
  \citenamefont {Ruloff}, \citenamefont {Wagner},\ and\ \citenamefont
  {Friedrich}}]{Klindt:2017}%
  \BibitemOpen
  \bibfield  {author} {\bibinfo {author} {\bibfnamefont {G.~S.}\ \bibnamefont
  {Klindt}}, \bibinfo {author} {\bibfnamefont {C.}~\bibnamefont {Ruloff}},
  \bibinfo {author} {\bibfnamefont {C.}~\bibnamefont {Wagner}},\ and\ \bibinfo
  {author} {\bibfnamefont {B.~M.}\ \bibnamefont {Friedrich}},\ }\bibfield
  {title} {\bibinfo {title} {In-phase and anti-phase flagellar synchronization
  by waveform compliance and basal coupling},\ }\href@noop {} {\bibfield
  {journal} {\bibinfo  {journal} {New J. Phys.}\ }\textbf {\bibinfo {volume}
  {19}},\ \bibinfo {pages} {113052} (\bibinfo {year} {2017})}\BibitemShut
  {NoStop}%
\bibitem [{\citenamefont {Liu}\ \emph {et~al.}(2018)\citenamefont {Liu},
  \citenamefont {Claydon}, \citenamefont {Polin},\ and\ \citenamefont
  {Brumley}}]{Liu2018}%
  \BibitemOpen
  \bibfield  {author} {\bibinfo {author} {\bibfnamefont {Y.}~\bibnamefont
  {Liu}}, \bibinfo {author} {\bibfnamefont {R.}~\bibnamefont {Claydon}},
  \bibinfo {author} {\bibfnamefont {M.}~\bibnamefont {Polin}},\ and\ \bibinfo
  {author} {\bibfnamefont {D.~R.}\ \bibnamefont {Brumley}},\ }\bibfield
  {title} {\bibinfo {title} {{Transitions in synchronization states of model
  cilia through basal-connection coupling}},\ }\href@noop {} {\bibfield
  {journal} {\bibinfo  {journal} {J. R. Soc. Interface}\ }\textbf {\bibinfo
  {volume} {15}},\ \bibinfo {pages} {20180450} (\bibinfo {year}
  {2018})}\BibitemShut {NoStop}%
\bibitem [{\citenamefont {Hamilto}\ and\ \citenamefont
  {Cicuta}(2021)}]{Hamilton:2021}%
  \BibitemOpen
  \bibfield  {author} {\bibinfo {author} {\bibfnamefont {E.}~\bibnamefont
  {Hamilto}}\ and\ \bibinfo {author} {\bibfnamefont {P.}~\bibnamefont
  {Cicuta}},\ }\bibfield  {title} {\bibinfo {title} {Changes in geometrical
  aspects of a simple model of cilia synchronization control the dynamical
  state, a possible mechanism for switching of swimming gaits in
  microswimmers},\ }\href@noop {} {\bibfield  {journal} {\bibinfo  {journal}
  {PLoS ONE}\ }\textbf {\bibinfo {volume} {16}},\ \bibinfo {pages} {e0249060}
  (\bibinfo {year} {2021})}\BibitemShut {NoStop}%
\bibitem [{\citenamefont {Chakrabarti}\ \emph {et~al.}(2022)\citenamefont
  {Chakrabarti}, \citenamefont {Fürthauer},\ and\ \citenamefont
  {Shelley}}]{Chakrabarti:2022}%
  \BibitemOpen
  \bibfield  {author} {\bibinfo {author} {\bibfnamefont {B.}~\bibnamefont
  {Chakrabarti}}, \bibinfo {author} {\bibfnamefont {S.}~\bibnamefont
  {Fürthauer}},\ and\ \bibinfo {author} {\bibfnamefont {M.~J.}\ \bibnamefont
  {Shelley}},\ }\bibfield  {title} {\bibinfo {title} {A multiscale biophysical
  model gives quantized metachronal waves in a lattice of beating cilia},\
  }\href {https://doi.org/10.1073/pnas.2113539119} {\bibfield  {journal}
  {\bibinfo  {journal} {Proceedings of the National Academy of Sciences}\
  }\textbf {\bibinfo {volume} {119}},\ \bibinfo {pages} {e2113539119} (\bibinfo
  {year} {2022})}\BibitemShut {NoStop}%
\bibitem [{\citenamefont {Uchida}\ and\ \citenamefont
  {Golestanian}(2010)}]{Uchida2010}%
  \BibitemOpen
  \bibfield  {author} {\bibinfo {author} {\bibfnamefont {N.}~\bibnamefont
  {Uchida}}\ and\ \bibinfo {author} {\bibfnamefont {R.}~\bibnamefont
  {Golestanian}},\ }\bibfield  {title} {\bibinfo {title} {{Synchronization and
  Collective Dynamics in a Carpet of Microfluidic Rotors}},\ }\href@noop {}
  {\bibfield  {journal} {\bibinfo  {journal} {Physical Review Letters}\
  }\textbf {\bibinfo {volume} {104}},\ \bibinfo {pages} {178103} (\bibinfo
  {year} {2010})}\BibitemShut {NoStop}%
\bibitem [{\citenamefont {Blake}(1971)}]{Blake1971}%
  \BibitemOpen
  \bibfield  {author} {\bibinfo {author} {\bibfnamefont {J.~R.}\ \bibnamefont
  {Blake}},\ }\bibfield  {title} {\bibinfo {title} {{A note on the image system
  for a stokeslet in a no-slip boundary}},\ }\href@noop {} {\bibfield
  {journal} {\bibinfo  {journal} {Mathematical Proceedings of the Cambridge
  Philosophical Society}\ }\textbf {\bibinfo {volume} {70}},\ \bibinfo {pages}
  {303} (\bibinfo {year} {1971})}\BibitemShut {NoStop}%
\bibitem [{\citenamefont {Gueron}\ \emph {et~al.}(1997)\citenamefont {Gueron},
  \citenamefont {Levit-Gurevich}, \citenamefont {Liron},\ and\ \citenamefont
  {Blum}}]{Gueron1997}%
  \BibitemOpen
  \bibfield  {author} {\bibinfo {author} {\bibfnamefont {S.}~\bibnamefont
  {Gueron}}, \bibinfo {author} {\bibfnamefont {K.}~\bibnamefont
  {Levit-Gurevich}}, \bibinfo {author} {\bibfnamefont {N.}~\bibnamefont
  {Liron}},\ and\ \bibinfo {author} {\bibfnamefont {J.~J.}\ \bibnamefont
  {Blum}},\ }\bibfield  {title} {\bibinfo {title} {{Cilia internal mechanism
  and metachronal coordination as the result of hydrodynamical coupling}},\
  }\href@noop {} {\bibfield  {journal} {\bibinfo  {journal} {Proceedings of the
  National Academy of Sciences}\ }\textbf {\bibinfo {volume} {94}},\ \bibinfo
  {pages} {6001} (\bibinfo {year} {1997})}\BibitemShut {NoStop}%
\bibitem [{\citenamefont {Ding}\ \emph {et~al.}(2014)\citenamefont {Ding},
  \citenamefont {Nawroth}, \citenamefont {McFall-Ngai},\ and\ \citenamefont
  {Kanso}}]{Ding2014}%
  \BibitemOpen
  \bibfield  {author} {\bibinfo {author} {\bibfnamefont {Y.}~\bibnamefont
  {Ding}}, \bibinfo {author} {\bibfnamefont {J.~C.}\ \bibnamefont {Nawroth}},
  \bibinfo {author} {\bibfnamefont {M.~J.}\ \bibnamefont {McFall-Ngai}},\ and\
  \bibinfo {author} {\bibfnamefont {E.}~\bibnamefont {Kanso}},\ }\bibfield
  {title} {\bibinfo {title} {{Mixing and transport by ciliary carpets: a
  numerical study}},\ }\href@noop {} {\bibfield  {journal} {\bibinfo  {journal}
  {Journal of Fluid Mechanics}\ }\textbf {\bibinfo {volume} {743}},\ \bibinfo
  {pages} {124} (\bibinfo {year} {2014})}\BibitemShut {NoStop}%
\bibitem [{\citenamefont {Schoeller}\ \emph {et~al.}(2021)\citenamefont
  {Schoeller}, \citenamefont {Townsend}, \citenamefont {Westwood},\ and\
  \citenamefont {Keaveny}}]{Schoeller:2021}%
  \BibitemOpen
  \bibfield  {author} {\bibinfo {author} {\bibfnamefont {S.~F.}\ \bibnamefont
  {Schoeller}}, \bibinfo {author} {\bibfnamefont {A.~K.}\ \bibnamefont
  {Townsend}}, \bibinfo {author} {\bibfnamefont {T.~A.}\ \bibnamefont
  {Westwood}},\ and\ \bibinfo {author} {\bibfnamefont {E.~E.}\ \bibnamefont
  {Keaveny}},\ }\bibfield  {title} {\bibinfo {title} {Methods for suspensions
  of passive and active filaments},\ }\href
  {https://doi.org/https://doi.org/10.1016/j.jcp.2020.109846} {\bibfield
  {journal} {\bibinfo  {journal} {Journal of Computational Physics}\ }\textbf
  {\bibinfo {volume} {424}},\ \bibinfo {pages} {109846} (\bibinfo {year}
  {2021})}\BibitemShut {NoStop}%
\bibitem [{\citenamefont {Martin}\ \emph {et~al.}(2019)\citenamefont {Martin},
  \citenamefont {Brunner},\ and\ \citenamefont {Deutsch}}]{Martin:2021}%
  \BibitemOpen
  \bibfield  {author} {\bibinfo {author} {\bibfnamefont {S.~E.}\ \bibnamefont
  {Martin}}, \bibinfo {author} {\bibfnamefont {M.~E.}\ \bibnamefont
  {Brunner}},\ and\ \bibinfo {author} {\bibfnamefont {J.~M.}\ \bibnamefont
  {Deutsch}},\ }\bibfield  {title} {\bibinfo {title} {Emergence of metachronal
  waves in active microtubule arrays},\ }\href
  {https://doi.org/10.1103/PhysRevFluids.4.103101} {\bibfield  {journal}
  {\bibinfo  {journal} {Phys. Rev. Fluids}\ }\textbf {\bibinfo {volume} {4}},\
  \bibinfo {pages} {103101} (\bibinfo {year} {2019})}\BibitemShut {NoStop}%
\bibitem [{\citenamefont {Vilfan}\ and\ \citenamefont
  {J{\"{u}}licher}(2006)}]{Vilfan2006}%
  \BibitemOpen
  \bibfield  {author} {\bibinfo {author} {\bibfnamefont {A.}~\bibnamefont
  {Vilfan}}\ and\ \bibinfo {author} {\bibfnamefont {F.}~\bibnamefont
  {J{\"{u}}licher}},\ }\bibfield  {title} {\bibinfo {title} {{Hydrodynamic Flow
  Patterns and Synchronization of Beating Cilia}},\ }\href@noop {} {\bibfield
  {journal} {\bibinfo  {journal} {Physical Review Letters}\ }\textbf {\bibinfo
  {volume} {96}},\ \bibinfo {pages} {058102} (\bibinfo {year}
  {2006})}\BibitemShut {NoStop}%
\bibitem [{\citenamefont {Qian}\ \emph {et~al.}(2009)\citenamefont {Qian},
  \citenamefont {Jiang}, \citenamefont {Gagnon}, \citenamefont {Breuer},\ and\
  \citenamefont {Powers}}]{Qian2009}%
  \BibitemOpen
  \bibfield  {author} {\bibinfo {author} {\bibfnamefont {B.}~\bibnamefont
  {Qian}}, \bibinfo {author} {\bibfnamefont {H.}~\bibnamefont {Jiang}},
  \bibinfo {author} {\bibfnamefont {D.~A.}\ \bibnamefont {Gagnon}}, \bibinfo
  {author} {\bibfnamefont {K.~S.}\ \bibnamefont {Breuer}},\ and\ \bibinfo
  {author} {\bibfnamefont {T.~R.}\ \bibnamefont {Powers}},\ }\bibfield  {title}
  {\bibinfo {title} {{Minimal model for synchronization induced by hydrodynamic
  interactions}},\ }\href@noop {} {\bibfield  {journal} {\bibinfo  {journal}
  {Physical Review E}\ }\textbf {\bibinfo {volume} {80}},\ \bibinfo {pages}
  {061919} (\bibinfo {year} {2009})}\BibitemShut {NoStop}%
\bibitem [{\citenamefont {Uchida}\ and\ \citenamefont
  {Golestanian}(2011)}]{Uchida2011}%
  \BibitemOpen
  \bibfield  {author} {\bibinfo {author} {\bibfnamefont {N.}~\bibnamefont
  {Uchida}}\ and\ \bibinfo {author} {\bibfnamefont {R.}~\bibnamefont
  {Golestanian}},\ }\bibfield  {title} {\bibinfo {title} {{Generic Conditions
  for Hydrodynamic Synchronization}},\ }\href@noop {} {\bibfield  {journal}
  {\bibinfo  {journal} {Physical Review Letters}\ }\textbf {\bibinfo {volume}
  {106}},\ \bibinfo {pages} {058104} (\bibinfo {year} {2011})}\BibitemShut
  {NoStop}%
\bibitem [{\citenamefont {Uchida}\ and\ \citenamefont
  {Golestanian}(2012)}]{Uchida2012}%
  \BibitemOpen
  \bibfield  {author} {\bibinfo {author} {\bibfnamefont {N.}~\bibnamefont
  {Uchida}}\ and\ \bibinfo {author} {\bibfnamefont {R.}~\bibnamefont
  {Golestanian}},\ }\bibfield  {title} {\bibinfo {title} {{Hydrodynamic
  synchronization between objects with cyclic rigid trajectories}},\
  }\href@noop {} {\bibfield  {journal} {\bibinfo  {journal} {The European
  Physical Journal E}\ }\textbf {\bibinfo {volume} {35}},\ \bibinfo {pages}
  {135} (\bibinfo {year} {2012})}\BibitemShut {NoStop}%
\bibitem [{\citenamefont {Guo}\ \emph {et~al.}(2018)\citenamefont {Guo},
  \citenamefont {Fauci}, \citenamefont {Shelley},\ and\ \citenamefont
  {Kanso}}]{Guo:2018}%
  \BibitemOpen
  \bibfield  {author} {\bibinfo {author} {\bibfnamefont {H.}~\bibnamefont
  {Guo}}, \bibinfo {author} {\bibfnamefont {L.}~\bibnamefont {Fauci}}, \bibinfo
  {author} {\bibfnamefont {M.}~\bibnamefont {Shelley}},\ and\ \bibinfo {author}
  {\bibfnamefont {E.}~\bibnamefont {Kanso}},\ }\bibfield  {title} {\bibinfo
  {title} {Bistability in the synchronization of actuated microfilaments},\
  }\href {https://doi.org/10.1017/jfm.2017.816} {\bibfield  {journal} {\bibinfo
   {journal} {Journal of Fluid Mechanics}\ }\textbf {\bibinfo {volume} {836}},\
  \bibinfo {pages} {304–323} (\bibinfo {year} {2018})}\BibitemShut {NoStop}%
\bibitem [{\citenamefont {Niedermayer}\ \emph {et~al.}(2008)\citenamefont
  {Niedermayer}, \citenamefont {Eckhardt},\ and\ \citenamefont
  {Lenz}}]{Niedermayer2008}%
  \BibitemOpen
  \bibfield  {author} {\bibinfo {author} {\bibfnamefont {T.}~\bibnamefont
  {Niedermayer}}, \bibinfo {author} {\bibfnamefont {B.}~\bibnamefont
  {Eckhardt}},\ and\ \bibinfo {author} {\bibfnamefont {P.}~\bibnamefont
  {Lenz}},\ }\bibfield  {title} {\bibinfo {title} {{Synchronization, phase
  locking, and metachronal wave formation in ciliary chains}},\ }\href@noop {}
  {\bibfield  {journal} {\bibinfo  {journal} {Chaos: An Interdisciplinary
  Journal of Nonlinear Science}\ }\textbf {\bibinfo {volume} {18}},\ \bibinfo
  {pages} {037128} (\bibinfo {year} {2008})}\BibitemShut {NoStop}%
\bibitem [{\citenamefont {Hickey}\ \emph {et~al.}(2023)\citenamefont {Hickey},
  \citenamefont {Golestanian},\ and\ \citenamefont {Vilfan}}]{Hickey:2023}%
  \BibitemOpen
  \bibfield  {author} {\bibinfo {author} {\bibfnamefont {D.~J.}\ \bibnamefont
  {Hickey}}, \bibinfo {author} {\bibfnamefont {R.}~\bibnamefont
  {Golestanian}},\ and\ \bibinfo {author} {\bibfnamefont {A.}~\bibnamefont
  {Vilfan}},\ }\href@noop {} {\bibinfo {title} {Nonreciprocal interactions give
  rise to fast cilium synchronisation in finite systems}} (\bibinfo {year}
  {2023}),\ \Eprint {https://arxiv.org/abs/2305.01077} {arXiv:2305.01077
  [cond-mat.soft]} \BibitemShut {NoStop}%
\bibitem [{\citenamefont {Hill}\ \emph {et~al.}(2010)\citenamefont {Hill},
  \citenamefont {Swaminathan}, \citenamefont {Estes}, \citenamefont {Cribb},
  \citenamefont {O'Brien}, \citenamefont {Davis},\ and\ \citenamefont
  {Superfine}}]{Hill:2010}%
  \BibitemOpen
  \bibfield  {author} {\bibinfo {author} {\bibfnamefont {D.~B.}\ \bibnamefont
  {Hill}}, \bibinfo {author} {\bibfnamefont {V.}~\bibnamefont {Swaminathan}},
  \bibinfo {author} {\bibfnamefont {A.}~\bibnamefont {Estes}}, \bibinfo
  {author} {\bibfnamefont {J.}~\bibnamefont {Cribb}}, \bibinfo {author}
  {\bibfnamefont {E.~T.}\ \bibnamefont {O'Brien}}, \bibinfo {author}
  {\bibfnamefont {C.~W.}\ \bibnamefont {Davis}},\ and\ \bibinfo {author}
  {\bibfnamefont {R.}~\bibnamefont {Superfine}},\ }\bibfield  {title} {\bibinfo
  {title} {Force generation and dynamics of individual cilia under external
  loading},\ }\href@noop {} {\bibfield  {journal} {\bibinfo  {journal}
  {Biophys. J.}\ }\textbf {\bibinfo {volume} {98}},\ \bibinfo {pages} {57}
  (\bibinfo {year} {2010})}\BibitemShut {NoStop}%
\bibitem [{\citenamefont {Sanderson}\ and\ \citenamefont
  {Sleigh}(1981)}]{Sanderson:1981}%
  \BibitemOpen
  \bibfield  {author} {\bibinfo {author} {\bibfnamefont {M.~J.}\ \bibnamefont
  {Sanderson}}\ and\ \bibinfo {author} {\bibfnamefont {M.~A.}\ \bibnamefont
  {Sleigh}},\ }\bibfield  {title} {\bibinfo {title} {{Ciliary activity of
  cultured rabbit tracheal epithelium: beat pattern and metachrony}},\ }\href
  {https://doi.org/10.1242/jcs.47.1.331} {\bibfield  {journal} {\bibinfo
  {journal} {Journal of Cell Science}\ }\textbf {\bibinfo {volume} {47}},\
  \bibinfo {pages} {331} (\bibinfo {year} {1981})}\BibitemShut {NoStop}%
\bibitem [{\citenamefont {Gauger}\ \emph {et~al.}(2009)\citenamefont {Gauger},
  \citenamefont {Downton},\ and\ \citenamefont {Stark}}]{Gauger2009}%
  \BibitemOpen
  \bibfield  {author} {\bibinfo {author} {\bibfnamefont {E.~M.}\ \bibnamefont
  {Gauger}}, \bibinfo {author} {\bibfnamefont {M.~T.}\ \bibnamefont
  {Downton}},\ and\ \bibinfo {author} {\bibfnamefont {H.}~\bibnamefont
  {Stark}},\ }\bibfield  {title} {\bibinfo {title} {{Fluid transport at low
  Reynolds number with magnetically actuated artificial cilia}},\ }\href@noop
  {} {\bibfield  {journal} {\bibinfo  {journal} {The European Physical Journal
  E}\ }\textbf {\bibinfo {volume} {28}},\ \bibinfo {pages} {231} (\bibinfo
  {year} {2009})}\BibitemShut {NoStop}%
\bibitem [{\citenamefont {Khaderi}\ \emph {et~al.}(2011)\citenamefont
  {Khaderi}, \citenamefont {den Toonder},\ and\ \citenamefont
  {Onck}}]{Khaderi2011}%
  \BibitemOpen
  \bibfield  {author} {\bibinfo {author} {\bibfnamefont {S.~N.}\ \bibnamefont
  {Khaderi}}, \bibinfo {author} {\bibfnamefont {J.~M.~J.}\ \bibnamefont {den
  Toonder}},\ and\ \bibinfo {author} {\bibfnamefont {P.~R.}\ \bibnamefont
  {Onck}},\ }\bibfield  {title} {\bibinfo {title} {{Microfluidic propulsion by
  the metachronal beating of magnetic artificial cilia: a numerical
  analysis}},\ }\href@noop {} {\bibfield  {journal} {\bibinfo  {journal}
  {Journal of Fluid Mechanics}\ }\textbf {\bibinfo {volume} {688}},\ \bibinfo
  {pages} {44} (\bibinfo {year} {2011})}\BibitemShut {NoStop}%
\bibitem [{\citenamefont {Chateau}\ \emph {et~al.}(2019)\citenamefont
  {Chateau}, \citenamefont {Favier}, \citenamefont {Poncet},\ and\
  \citenamefont {D'Ortona}}]{Chateau:2019}%
  \BibitemOpen
  \bibfield  {author} {\bibinfo {author} {\bibfnamefont {S.}~\bibnamefont
  {Chateau}}, \bibinfo {author} {\bibfnamefont {J.}~\bibnamefont {Favier}},
  \bibinfo {author} {\bibfnamefont {S.}~\bibnamefont {Poncet}},\ and\ \bibinfo
  {author} {\bibfnamefont {U.}~\bibnamefont {D'Ortona}},\ }\bibfield  {title}
  {\bibinfo {title} {Why antiplectic metachronal cilia waves are optimal to
  transport bronchial mucus},\ }\href
  {https://doi.org/10.1103/PhysRevE.100.042405} {\bibfield  {journal} {\bibinfo
   {journal} {Phys. Rev. E}\ }\textbf {\bibinfo {volume} {100}},\ \bibinfo
  {pages} {042405} (\bibinfo {year} {2019})}\BibitemShut {NoStop}%
\bibitem [{\citenamefont {Guo}\ \emph {et~al.}(2014)\citenamefont {Guo},
  \citenamefont {Nawroth}, \citenamefont {Ding},\ and\ \citenamefont
  {Kanso}}]{Guo:2014}%
  \BibitemOpen
  \bibfield  {author} {\bibinfo {author} {\bibfnamefont {H.}~\bibnamefont
  {Guo}}, \bibinfo {author} {\bibfnamefont {J.}~\bibnamefont {Nawroth}},
  \bibinfo {author} {\bibfnamefont {Y.}~\bibnamefont {Ding}},\ and\ \bibinfo
  {author} {\bibfnamefont {E.}~\bibnamefont {Kanso}},\ }\bibfield  {title}
  {\bibinfo {title} {{Cilia beating patterns are not hydrodynamically
  optimal}},\ }\href {https://doi.org/10.1063/1.4894855} {\bibfield  {journal}
  {\bibinfo  {journal} {Physics of Fluids}\ }\textbf {\bibinfo {volume} {26}},\
  \bibinfo {pages} {091901} (\bibinfo {year} {2014})}\BibitemShut {NoStop}%
\end{thebibliography}

\begin{thebibliography}{1}

\bibitem{Blake1971}
J~R Blake.
\newblock {A note on the image system for a stokeslet in a no-slip boundary}.
\newblock {\em Mathematical Proceedings of the Cambridge Philosophical
  Society}, 70(02):303, 1971.

\bibitem{Meng:2021}
Fanlong Meng, Rachel~R. Bennett, Nariya Uchida, and Ramin Golestanian.
\newblock Conditions for metachronal coordination in arrays of model cilia.
\newblock {\em Proc. Natl. Acad. Sci.}, 118:e2102828118, 2021.

\end{thebibliography}
\end{document}